\documentclass{aastex63}
\usepackage{natbib}
\usepackage{graphicx}
\usepackage{amsmath}
\usepackage{acro}
\acsetup{
  single         = false,
}

\DeclareAcronym{EVE}{
  short = EVE,
  long = \textit{Extreme Ultraviolet Variability Experiment},
  cite = {Woods:2012}}

\DeclareAcronym{AIA}{
  short = AIA,
  long = \textit{Atmospheric Imaging Assembly},
  cite = {Lemen:2012}}

\DeclareAcronym{HMI}{
  short = HMI,
  long = \textit{Helioseismic Magnetic Imager},
  cite = {Scherrer:2012}}
  
\DeclareAcronym{SDO}{
  short = SDO,
  long = \textit{Solar Dynamics Observatory},
  cite = {Pesnell:2012}}

\DeclareAcronym{STEREO}{
  short = STEREO,
  long = \textit{Solar Terrestrial Relations Observatory},
  cite = {Kaiser:2008}}

\DeclareAcronym{STEREO-A}{
  short = STEREO,
  long = \textit{Solar Terrestrial Relations Observatory (Ahead)},
  cite = {Kaiser:2008}}
  
\DeclareAcronym{STEREO-B}{
  short = STEREO,
  long = \textit{Solar Terrestrial Relations Observatory (Behind)},
  cite = {Kaiser:2008}}

\DeclareAcronym{SECCHI}{
  short = SECCHI,
  long = \textit{Sun Earth Connection Coronal and Heliospheric Investigation},
  cite = {Howard:2008}}
  
\DeclareAcronym{SCIP}{
  short = SCIP,
  long = Sun Centered Imaging Package}
  
\DeclareAcronym{MGN}{
  short = MGN,
  long = multi-scale Gaussian normalisation,
  cite = {Morgan:2014}}

\DeclareAcronym{EUVI}{
  short = EUVI,
  long = EUVI,
  long = \textit{Extreme Ultraviolet Imager},
  cite = {Wuelser:2004}}

\DeclareAcronym{CME}{
  short = CME,
  short-plural-form = CMEs,
  long = coronal mass ejection,
  long-plural-form = coronal mass ejections}
  
\DeclareAcronym{PIL}{
  short = PIL,
  short-plural-form = PILs,
  long = polarity inversion line,
  long-plural-form = polarity inversion lines}
  
\DeclareAcronym{LOS}{
  short = LOS,
  short-plural-form = LOS,
  long = line of sight,
  long-plural-form = lines of sight}
  
\DeclareAcronym{FOV}{
  short = FOV,
  short-plural-form = FOV,
  long = field of view,
  long-plural-form = fields of view}
  
\DeclareAcronym{EUV}{
  short = EUV,
  short-plural-form = EUV,
  long = extreme ultraviolet,
  long-plural-form = extreme ultraviolet}
  
\DeclareAcronym{GONG}{
  short = GONG,
  long = \textit{Global Oscillations Network Group}}
  
\DeclareAcronym{DST}{
  short = DST,
  long = \textit{Dunn Solar Telescope}}

\DeclareAcronym{IBIS}{
  short = IBIS,
  long = \textit{Interferometric Bidimensional Spectropolarimeter},
  cite = {Cavallini:2006,Reardon:2008}}

\DeclareAcronym{FIRS}{
  short = FIRS,
  long = \textit{Facility Infrared Spectropolarimeter},
  cite = {Jaeggli:2011}}
  
\DeclareAcronym{EIT}{
  short = EIT,
  long = \textit{Extreme Ultraviolet Imaging Telescope},
  cite = {Delaboudiniere:1995}}
 
\DeclareAcronym{ROSA}{
  short = ROSA,
  long = \textit{Rapid Oscillations in the Solar Atmosphere},
  cite = {Jess:2010}}
  
\DeclareAcronym{SPINOR}{
  short = SPINOR,
  long = \textit{Spectro-Polarimeter for Infrared and Optical Regions},
  cite = {Socas:2006}}  
 
\DeclareAcronym{SOHO}{
  short = SOHO,
  long = \textit{Solar and Heliospheric Observatory},
  cite = {Domingo:1995}}
 
\DeclareAcronym{MDI}{
  short = MDI,
  long = \textit{Michelson Doppler Imager},
  cite = {Scherrer:1995}}

\DeclareAcronym{NMSU}{
  short = NMSU,
  long = New Mexico State University}

\DeclareAcronym{NSO}{
  short = NSO,
  long = \textit{National Solar Observatory}}
  
\DeclareAcronym{TAC}{
  short = TAC,
  long = time allocation committee}
  
\DeclareAcronym{DEM}{
  short = DEM,
  short-plural-form = DEMs,
  long = Differential Emission Measure,
  long-plural-form = Differential Emission Measures}
  
\DeclareAcronym{PI}{
  short = PI,
  short-plural-form = PI,
  long = Principle Investigator,
  long-plural-form = Principle Investigators}  
  
\DeclareAcronym{IDL}{
  short = IDL,
  long = Interactive Data Language}
  
\DeclareAcronym{bb}{
  short = bb,
  long = broadband}  
  
\DeclareAcronym{nb}{
  short = nb,
  long = narrowband}    
  
\DeclareAcronym{IPM}{
  short = IPM,
  long = interplanetary medium}  
  
\DeclareAcronym{AO}{
  short = AO,
  long = adaptive optics,
  cite = {Rimmele:2004}} 
  
\DeclareAcronym{IR}{
  short = IR,
  long = infrared}  

\DeclareAcronym{halpha}{
  short = H$\alpha$,
  long = H$\alpha$}

\DeclareAcronym{FPI}{
  short = FPI,
  long = Fabry-P\'erot}
  
\DeclareAcronym{DWDM}{
  short = DWDM,
  long = dense wavelength division multiplexing}  

\DeclareAcronym{CCD}{
  short = CCD,
  long = charge-coupled device}   
  
\DeclareAcronym{HI}{
  short = HI,
  long = Heliospheric Investigation}   
  
\DeclareAcronym{ROI}{
  short = ROI,
  long = region-of-interest}    
  
\DeclareAcronym{POV}{
  short = POV,
  long = point-of-view}  
  
\DeclareAcronym{MHS}{
  short = MHS,
  long = magnetohydrostatic}
  
\DeclareAcronym{MHD}{
  short = MHD,
  long = magnetohydrodynamic}  
  
\DeclareAcronym{DOT}{
  short = DOT,
  long = \textit{Dutch Open Telescope},
  cite = {Rutten:1997}}  
  
\DeclareAcronym{BBSO}{
  short = BBSO,
  long = \textit{Big Bear Solar Observatory}}
  
\DeclareAcronym{NST}{
  short = NST,
  long = \textit{New Solar Telescope},
  cite = {Goode:2012}}

\DeclareAcronym{SOT}{
  short = SOT,
  long = \textit{Solar Optical Telescope},
  cite = {Tsuneta:2008}}
  
\DeclareAcronym{hinode}{
  short = Hinode,
  long = Hinode,
  cite = {Kosugi:2007}}
  
\DeclareAcronym{GST}{
  short = GST,
  long = \textit{Goode Solar Telescope}}  
  
\DeclareAcronym{SST}{
  short = SST,
  long = \textit{Swedish 1-m Solar Telescope},
  cite = {Scharmer:2002}}  
  
\DeclareAcronym{TRACE}{
  short = TRACE,
  long = \textit{Transition Region and Coronal Explorer},
  cite = {Handy:1999}}  
  
\DeclareAcronym{PCTR}{
  short = PCTR,
  long = \textit{prominence-corona-transition-region}}
  
\DeclareAcronym{DKIST}{
  short = DKIST,
  long = Daniel K. Inouye Solar Telescope}
  
\DeclareAcronym{RTI}{
  short = RTI,
  long = Rayleigh-Taylor instability}
  
\DeclareAcronym{SSW}{
  short = SSW,
  long = SolarSoftWare,
  cite = {Freeland:1998}}    
  
\DeclareAcronym{LTE}{
  short = LTE,
  long = local thermodynamic equilibrium}
  
\DeclareAcronym{NLTE}{
  short = NLTE,
  long = non-local thermodynamic equilibrium}
  
\DeclareAcronym{FTS}{
  short = FTS,
  long = Fourier Transform Spectrometer,
  cite = {Kurucz:1984}}
  
\DeclareAcronym{RTE}{
  short = RTE,
  long = radiative transfer equation}
  
\DeclareAcronym{CRD}{
  short = CRD,
  long = complete redistribution}
  
\DeclareAcronym{PRD}{
  short = PRD,
  long = partial redistribution}
  
\DeclareAcronym{BCM}{
	short = BCM,
	long = Beckers' cloud model,
	cite={Beckers:1964}}
	
\DeclareAcronym{HSRA}{
	short = HSRA,
	long = Harvard Smithsonian Reference Atmosphere,
	cite={Gingerich:1971}}
	
\DeclareAcronym{NICOLE}{
	short = NICOLE,
	long = Non-LTE Inversion COde using the Lorien Engine,
	cite={Socasnavarro:2015}}
	
\DeclareAcronym{SIR}{
	short = SIR,
	long = Stokes Inversion based on Response functions,
	cite={Ruizcobo:1992}}
	
\DeclareAcronym{HAZEL}{
	short = HAZEL,
	long = Hanle and Zeeman Light,
	cite={AsensioRamos:2008}}
	
\DeclareAcronym{BPSS}{
	short = BPSS,
	long = bald patch separatrix surface,
	cite={Bungey:1996}}
	
\DeclareAcronym{EIS}{
	short = EIS,
	long = \textit{EUV Imaging Spectrometer},
	cite={Culhane:2007}}

\DeclareAcronym{FWHM}{
  short = FWHM,
  long = full width at half maximum}

\DeclareAcronym{AMRVAC}{
	short = MPI-AMRVAC,
	long = \textit{Adaptive Mesh Refinement Versatile Advection Code},
	cite = {Keppens:2012,Porth:2014,Xia:2018}}

\DeclareAcronym{AMR}{
	short = AMR,
	long = adaptive mesh refinement}

\DeclareAcronym{CCI}{
	short = CCI,
	long = Convective Continuum Instability}

\DeclareAcronym{BV}{
	short = BV,
	long = Brunt-V\"ais\"al\"a}

\DeclareAcronym{VTT}{
    short = VTT,
    long = \textit{Vacuum-Tower Telescope},
    cite = {Vonderluhe:1998}}

\DeclareAcronym{Chrotel}{
    short = ChroTel,
    long = \textit{Chromospheric Telescope},
    cite = {Kentischer:2008}}

\DeclareAcronym{COR2}{
    short = COR2,
    long = \textit{Coronagraph 2},
    cite = {Howard:2008}}

\DeclareAcronym{PSP}{
    short = PSP,
    long = \textit{Parker Solar Probe},
    cite = {Fox:2016}}

\DeclareAcronym{SO}{
    short = SO,
    long = \textit{Solar Orbiter},
    cite = {Muller:2020}}
\graphicspath{{./}{figures/}}
\shortauthors{Wang et al.}

\begin{document}
\title{Velocities of an Erupting Filament}

\author[0000-0001-5589-0416]{Shuo Wang}
\affil{Department of Astronomy, New Mexico State University, P.O. Box 30001, MSC 4500, Las Cruces, NM 88003, USA}
\footnote{DKIST Ambassador}

\author[0000-0002-8975-812X]{Jack M. Jenkins}
\affiliation{Centre for mathematical Plasma Astrophysics, Department of Mathematics, KU Leuven, Celestijnenlaan 200B, B-3001 Leuven, Belgium}

\author[0000-0002-5547-9683]{Karin Muglach}
\affiliation{Catholic University of America, Washington, DC 20064, USA}
\affiliation{NASA Goddard Space Flight Center, Greenbelt, MD 20771, USA}

\author[0000-0001-7764-6895]{Valentin Martinez Pillet}
\affiliation{National Solar Observatory, 3665 Discovery Drive, Boulder, CO 80303, USA}

\author[0000-0001-7706-4158]{Christian Beck}
\affiliation{National Solar Observatory, 3665 Discovery Drive, Boulder, CO 80303, USA}

\author[0000-0003-3137-0277]{David M. Long}
\affiliation{Mullard Space Science Laboratory, University College London, Holmbury St. Mary, Dorking, Surrey, RH5 6NT, UK}

\author[0000-0002-9308-3639]{Debi Prasad Choudhary}
\affiliation{Department of Physics and Astronomy, California State University Northridge, Northridge, CA 91330-8268, USA}

\author[0000-0003-1493-101X]{James McAteer}
\affil{Department of Astronomy, New Mexico State University, P.O. Box 30001, MSC 4500, Las Cruces, NM 88003, USA}
\affil{Sunspot Solar Observatory, Sunspot, NM 88349, USA}

\correspondingauthor{Shuo Wang}
\email{shuowang@nmsu.edu}

\begin{abstract}
Solar filaments exist as stable structures for extended periods of time before many of them form the core of a \ac{CME}. We examine the properties of an erupting filament on 2017 May 29--30 with high-resolution He~{\sc i}~10830~{\AA} and \acs{halpha} spectra from the Dunn Solar Telescope, full-disk Dopplergrams of He~{\sc i}~10830~{\AA} from the Chromospheric Telescope, and EUV and coronograph data from SDO and STEREO. Pre-eruption line-of-sight velocities from an inversion of He~{\sc i} with the HAZEL code exhibit coherent patches of 5~Mm extent that indicate counter-streaming and/or buoyant behavior. During the eruption, individual, aligned threads appear in the He~{\sc i} velocity maps. The distribution of velocities evolves from Gaussian to strongly asymmetric. The maximal optical depth of  He~{\sc i}~10830~{\AA} decreased from $\tau = 1.75$ to 0.25, the temperature increased by 13 kK, and the average speed and width of the filament increased from 0 to 25~km~s$^{-1}$ and 10 to 20~Mm, respectively. All data sources agree that the filament rose with an exponential acceleration reaching 7.4~m~s$^{-2}$  that increased to a final velocity of 430~km~s$^{-1}$ at 22:24 UT; a \ac{CME} was associated with this filament eruption. The properties during the eruption favor a kink/torus instability, which requires the existence of a flux rope. We conclude that full-disk chromospheric Dopplergrams can be used to trace the initial phase of on-disk filament eruptions in real-time, which might potentially be useful for  modelling the source of any subsequent CMEs.
\end{abstract}

\keywords{Sun: filaments, prominences --- Sun: infrared --- Sun: activity --- Sun: coronal mass ejections (CMEs)}

\section{Introduction} \label{sec:intro}
Solar filaments are condensations of relatively cool plasma suspended at coronal heights within the solar atmosphere. When projected against the solar disk, their cool properties render them in absorption whereas their off-limb counterpart, prominences, appear bright against the dark background of space \citep[][]{Labrosse:2010, Mackay:2010, Vial:2015, Gibson:2018}. With lengths of several tens to hundreds of Mm, and heights and widths of only a few to several ten Mms they are amongst the longest structures in the solar atmosphere, often appearing as elongated channels of chromospheric plasma that snake across the solar disk.

Filaments and prominences exist within the solar atmosphere for periods ranging from a few hours to a few months. The shortest-lived  samples tend to be ejected from the solar atmosphere during eruptions, many of which are cotemporal with flares and \acp{CME} \citep[][]{Green:2018}. The longer-lived  samples are relatively slow to evolve and their end-of-life dynamics can vary from weak, partial eruptions \citep[\textit{e.g.},][]{Choudhary:2003} to large interplanetary \acp{CME} \citep[\textit{e.g.},][]{Wood:2016}, thermal disparitions brusques \citep{1983SoPh...88..241S},  or a complete decay of the structure as the topology of the host magnetic field evolves in such a way as to no longer provide support for the filament material against gravity \citep[\textit{e.g.},][]{Jing:2003}.

Despite the slow global evolution of the latter cases, they often exhibit a highly dynamic structure at smaller scales.
The wide range of small Mm-scale, presumably thermodynamically-driven plasma evolutions within stable filaments and prominences have been studied for many decades \citep[][]{Leroy:1989,Engvold:1990}. \citet{Zirker:1998} reported on counter-streaming \textit{i.e.}, oppositely-oriented flows within filaments with velocities as high as 20~km~s$^{-1}$ \citep[see also][]{Litvinenko:1999, Wang:1999b, Alexander:2013, Ahn:2010, Diercke:2018}. Similar observational signatures have also been interpreted as buoyant or gravitational flows with velocities of the order of 10~km~s$^{-1}$ \citep[\textit{e.g.},][]{Berger:2011, Hillier:2018}, or small-scale oscillations in the host magnetic field itself \citep[\textit{e.g.},][]{Lin:2007}. On intermediate $\approx$~10~Mm scales, the motions of plasma within filaments and prominences have historically been related to the evolution of the host magnetic field rather than a thermodynamic driver \citep[see the reviews by][]{Tripathi:2009,Arregui:2018}. For example, the particularly large amplitude oscillations are reserved for those filaments in the vicinity of a flare, wherein the filaments are subjected to the expanding magnetic pressure bubble of the nearby eruption. The amplitudes of such dynamics are also significantly larger than those at the smaller scales, with velocities and displacements in the region of 30\,--\,100~km~s$^{-1}$ and 110~Mm, respectively \citep[\textit{e.g.},][]{Luna:2012, Luna:2014, Liakh:2020}. More recently, similarly large-scale and correlated mass motions occurring in the lead-up to a filament eruption have been added to the conditions for global flux rope stability \citep[\textit{e.g.},][]{Bi:2014, Reva:2017, Jenkins:2018, Jenkins:2019, 2020ApJ...898...34F}, alongside the more commonly-considered stability conditions \citep[\textit{e.g.}, torus/kink instability, breakout reconnection, tether cutting, etc.;][]{Antiochos:1999,Moore:2001,Torok:2005,Kliem:2006}.

 Unlike the motions of plasma within stable on-disk filaments, the study of the behaviour of filament plasma within the early stages of an on-disk eruption is far less common due to the scarcity of spectral observations for such events, although some examples do exist \citep[\textit{e.g.},][]{Muglach:1997,Penn:2000,Sasso:2011,Sasso:2014,Doyle:2019}. Once the eruptive filaments and prominences have propagated further out into the upper corona, the motion of the associated plasma is routinely tracked using automated algorithms \citep[\textit{e.g.},][]{Byrne:2015}, although many of these methods focus more on the white-light \ac{CME} component than the embedded filament. Where possible, some authors have previously aimed to obtain a more-complete picture (e.g., eruption trigger mechanism) by also including a more detailed analysis of the evolution of the associated filament \citep[\textit{e.g.}, mass evolution or the relationship between the 3D global shape and the inferred background decay index;][]{Seaton:2011,ReesCrockford:2020}.

Authors have previously obtained observations of eruptive filaments using the optically thin He~{\sc i}~10830~\AA\ or He D$_3$ lines \citep{LopezAriste:2015}, which enabled them to adopt the assumption of a Gaussian absorption profile in their inversion methods. In each case, the authors concluded that the measured spectral profile for He~{\sc i} yielded a satisfactory fit only if multiple summed Gaussians were employed, which indicates multiple velocity components within the line of sight of the erupting structure \citep[see specifically][]{Sasso:2014}. Velocities extracted from these fits range between 60\,--\,300~km~s$^{-1}$, comparable to the velocities of prominences in the plane-of-sky depending on the eruption stage \citep[\textit{e.g.},][]{McCauley:2015}. \citet{Doyle:2019} recently used similar assumptions to characterise the evolution of an erupting filament recorded in the more-readily observed optically thick \ac{halpha} line, also measuring speeds of $\approx$~60~km~s$^{-1}$. However, the study of \citet{Chae:2006} suggests such approximations applied to the analysis of optically-thick spectral lines are only first-order accurate.

Erupting filaments that lead to \acp{CME} are one of the main drivers of space weather near  Earth. A major goal of the solar community is to establish a  network of ground-based facilities that enables routine observations of the Sun at wavelengths that permit the extraction of crucial parameters for space weather modelling tools \citep{2019BAAS...51c.110M}. The two main parameters are the velocity to infer the travel time of a given eruption from the Sun to the Earth and the magnetic field orientation to ascertain whether an interaction between the associated \ac{CME} and the Earth's magnetosphere will be geo-effective \citep[][]{2018ApJ...864...18S,Owens:2020}. The aforementioned automated methods for extracting the velocities of CMEs are well-suited for those events where the \ac{CME} propagates close to the plane-of-sky and therefore the observed projected 2D \ac{CME} speed is close to the actual 3D speed of the \ac{CME} \citep[\textit{e.g.},][]{Byrne:2015}. On the other hand, these methods typically fail for those eruptions which have a significant component along the \ac{LOS}, \textit{i.e.}, towards the Earth.  This problem can of course be mitigated with suitable observations from an angle away from the Sun-Earth line, \textit{e.g.}, using instruments on board the \ac{STEREO} spacecraft \citep[][]{Owens:2020,Barnard:2020}. Furthermore, we are yet to routinely measure the magnetic field of the corona, including \acp{CME}, although some preliminary efforts have been made, \textit{e.g.}, \citet[][]{Bak-Steslicka:2013}. Fortunately, and as already indicated, eruptive filaments embedded within these \acp{CME} may prove to be ideal candidates for providing the initial velocity and magnetic field properties of their host eruptive structures \citep[\textit{e.g.},][]{Kuckein:2020,2020JSWSC..10...41H}. 

In \citet[][hereafter Paper I]{Wang:2020} we have already demonstrated that the magnetic field may be routinely extracted from erupting filaments observed in He~{\sc I}~10830~{\AA}.  Paper I focused primarily on the derivation of the magnetic field structure of the erupting filament that we will further study here, and found it to be consistent with a flux rope. The magnetic maps exhibited a large variation of field strengths, peaking above the 90th percentile value of 435~G, with average values of 24, 70, and 45~G during the eruption. At the same times, the magnetic field azimuth and inclination (to the vertical) were found to gradually increase from 48 to 54 degrees and decrease from 80 to 63 degrees, respectively.  \citet{Schwartz:2019} presented a \ac{NLTE} inversion study of \ac{halpha} plasma parameters for the pre-eruptive phase of the same filament on 2017 May 29, where they found a temperature range from 6--14 kK and non-thermal velocities from 4--9 km\,s$^{-1}$ over six different locations inside the filament. This paper is the continuation of Paper I with the objective to derive He~{\sc I}~10830~{\AA} plasma diagnostics from the further application of the \acf{HAZEL} code to measure \textit{e.g.}, the velocities within the erupting filament, which are complemented by a variety of measurements from other ground-based and space-based sources. In Section 2 we briefly describe our data sets. The methods for analysing spectral and image data are provided in Section 3. We present the results of the application of these methods to the spectral and imaging observations in Section 4. Section 5 and 6 give the discussion and conclusions, respectively.

\section{Observations} \label{sec:obs}

\begin{figure}
\centerline{\includegraphics[width=0.9\textwidth]{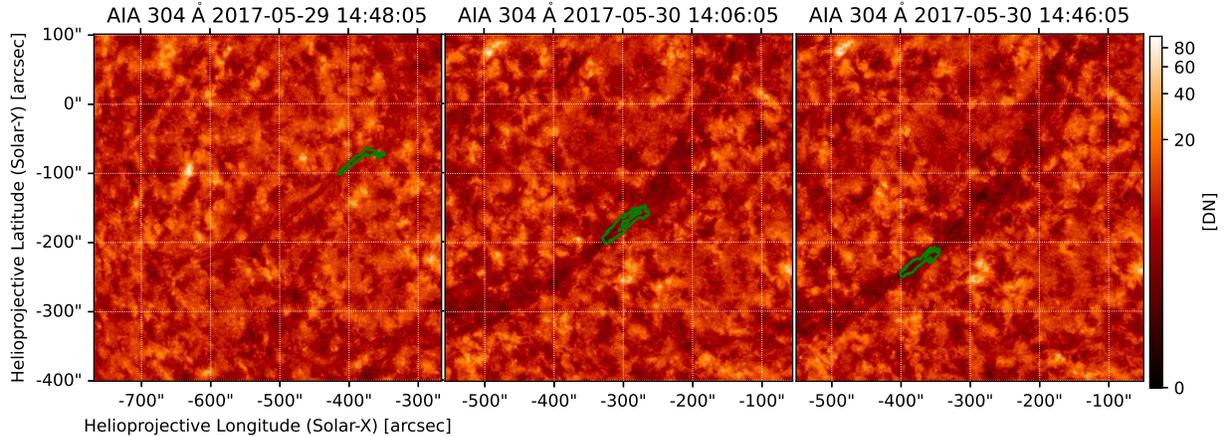}}
\caption{SDO/AIA 304~\AA\ maps on 2017 May 29 and 30. Contours mark the parts of filament observed by FIRS at 10830~\AA\ and IBIS at  \ac{halpha} 6563~\AA. The image intensity is in units of digital numbers (DN) and scaled logarithmically. These images show a subfield of the FOV of the animation of the prominence eruption that is available in the online material.  The animation covers the evolution at this part of the Sun between 00:06 UT and 18:56 UT on 2017 May 30 with a changing cadence of 2\,--\,10 minutes. The duration of the animation is seven seconds.
\label{fig:fig1}}
\end{figure}

\begin{figure}
\centerline{\includegraphics[width=0.85\textwidth]{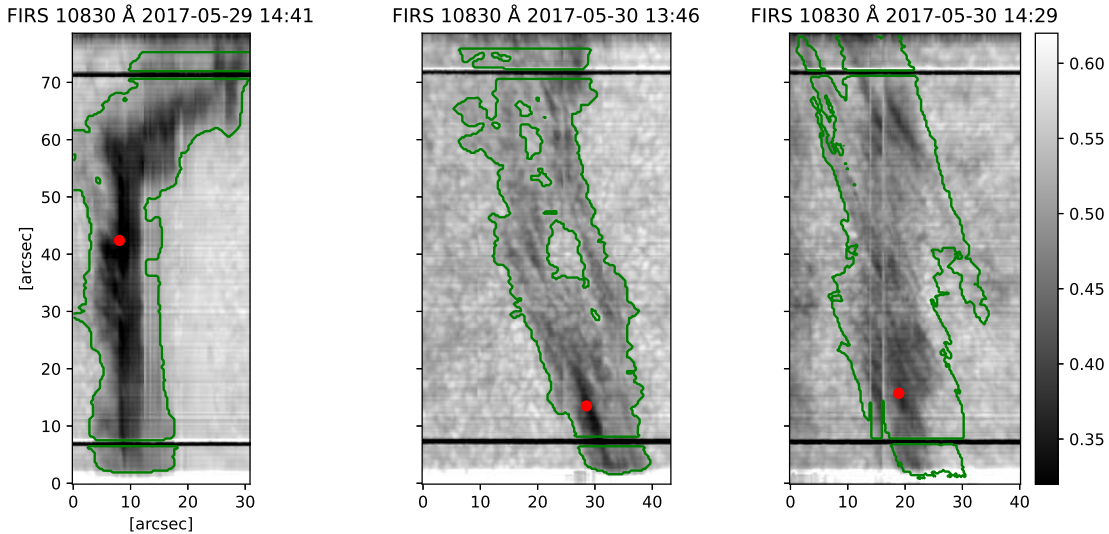}}
\caption{Line-core intensity map of He~{\sc I}~10830~{\AA} as observed by FIRS on 2017 May 29 and 30. Green contours show the borders of the regions of interest (ROIs) used for the inversion in Figure~\ref{fig:fig2}. Red points mark the positions of the pixels in the spine shown in Figure~\ref{fig:fig3}. The absorption by the He I line  decreases during the eruption leading to rising line-core intensities.  The two vertical bright stripes at $x\sim15^{\prime\prime}$ in the rightmost panel with the data on May 30 at 14:29 UT were caused by a temporarily loss of the lock point of the adaptive optics system.\label{fig:fig9}}
\end{figure}

From 23-29 May 2017, multiple Earth-positioned observatories recorded a long ($\approx$~660$''$) and stable quiescent filament stretching across the south-eastern quadrant of the solar surface. At approximately 12:00~UT on 30 May 2017, the filament erupted, for instance seen in  \ac{AIA} 304~{\AA} images, propagating to the south east (as projected on the solar disk from Earth view).  An animation of the 304~{\AA} observations is available in the online material. It shows the filament eruption on 2017 May 30 from 00:06~UT to 18:56~UT in AIA and \acf{EUVI} 304~{\AA} images in top and side view. Solid white curves in the animation indicate the location of the solar limb as seen by STEREO-A.

The \acf{FIRS} and \acf{IBIS} instruments installed at the \ac{DST} recorded this filament before and during the eruption on 2017 May 29 and 30, respectively. Over the  two observing days, the \ac{FIRS} instrument observed the He~{\sc i}~10830~\AA\ spectra and completed four full rasters across the width of the filament at 14:41 \& 15:07~UT on 29 May, and 13:46 \& 14:29~UT on 30 May. The position of these rasters relative to the entire filament is shown in Figure~\ref{fig:fig1} as the green contour overlaid on full-disk He~{\sc ii}~304~\AA\ observations provided by the \ac{AIA} instrument on board the \ac{SDO} (see also Figure~\ref{fig:fig9} for the associated contour definitions and explicit timestamps).  The telescope pointing at the DST covered three different parts of the filament, one part towards the northern end of the filament twice on May 29 and two different sections along the filament body on May 30. The \ac{FOV} for the \ac{IBIS} instrument was centred on the same location as the \ac{FIRS} \ac{FOV}. The \ac{IBIS} instrument observed during the same four time windows as \ac{FIRS}, continuously recording both \ac{halpha}~6562.8~\AA\ and Ca~{\sc ii}~8542~\AA\ intensity spectra at a cadence of 12~s. The Ca~{\sc ii}~8542~\AA\ spectra did not show a clear signature of this quiescent, high ($>> 10$\,Mm; Paper I) filament, especially during the eruption, as for instance also seen in \citet[][their Figure 6]{Beck:2018} for another quiescent filament . The off-center location of the filament also led to an inclined \ac{LOS}, which should reduce the opacity in the presumably more vertical structures closer to the photosphere that would be seen in Ca~{\sc ii}. Since the Ca~{\sc ii} spectra do not capture the filament body and only show traces of the filament foot points on May 29 one day prior to the eruption, we thus discarded them in the current investigation as they do not provide additional information on the conditions within the filament body above and beyond that provided by He~{\sc i}~10830~\AA. The orientation of the rotating coud\'e table was adjusted such that the slit of the \ac{FIRS} instrument was roughly aligned with the main axis of the filament, as can be seen in Figure~\ref{fig:fig9}. Descriptions of the full  setup for both the \ac{FIRS} and \ac{IBIS} instruments, including their data reduction, may be found in Paper I and \citet{Schwartz:2019}, respectively.

The \acf{Chrotel} was also observing during this period from 07:15 to 17:09 UT on 30 May 2017. The \ac{Chrotel} instrument observes the full-disk of the Sun with Lyot filters centred on Ca~{\sc ii}~K 3933, \ac{halpha} 6562.8, and He~{\sc i} 10830~\AA . The Lyot filter at 10830~\AA\ can be tuned to obtain Dopplergrams of the He~{\sc i} line \citep[][]{Bethge:2011}. Finally, at 12:00~UT on 30 May 2017, the  STEREO-A spacecraft was positioned approximately 136$^\circ$ behind the Earth in its orbit and viewed the eruption of the filament from the side with its  EUVI and \acf{COR2} instruments.

\section{Data Analysis}
\subsection{ He~{\sc i}~10830~{\AA} Inversion with Hazel}
The telluric H$_2$O line at 10832.108~\AA\ was first used to determine an accurate rest wavelength for the solar spectral lines. The HAZEL code was then used to invert the He~{\sc i}~10830~\AA\ intensity spectra from FIRS. The \ac{HAZEL} model assumes a slab with constant physical parameters at a fixed altitude above the solar surface.  The height of this slab was set to 15\,Mm on May 29, and to 33 and 79\,Mm on May 30 during the eruption as in Paper I. The \ac{ROI} for the inversion is indicated by the green contour in Figure~\ref{fig:fig9}. The full-resolution Stokes I data were used as the input  and the input magnetic field was set to zero. 

\begin{figure}
\centerline{\includegraphics[width=0.85\textwidth]{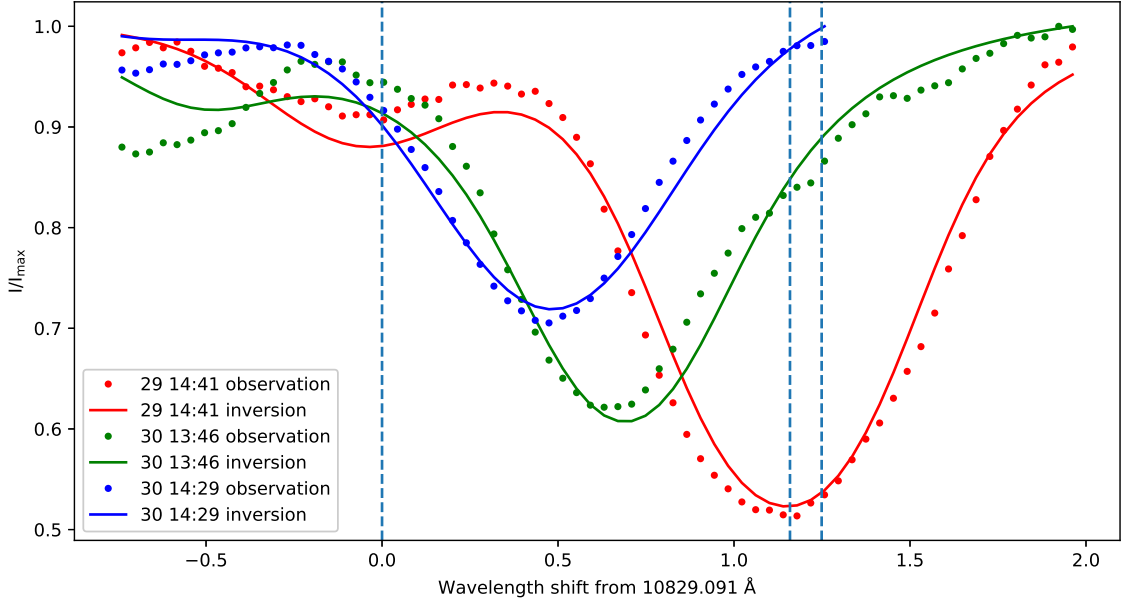}}
\caption{Profiles of pixels in the filament spine at different times.  The positions of the three pixels are marked with red dots in Figure~\ref{fig:fig9}. The blue dashed vertical lines mark the positions of the wavelengths for the He~{\sc i}~10830~\AA\ triplet at rest.
\label{fig:fig3}}
\end{figure}

Figure~\ref{fig:fig3} shows examples of the fitting wherein positions with a large line depth have been selected to show clear line profiles with small noise. Each of the three line profiles exhibits a single dominant component with symmetric line wings. The line profiles were observed at three different times and at three different locations in the filament as a consequence of the change in the position of the \ac{FOV} relative to the filament (Figure \ref{fig:fig1}). As time progresses, the LOS velocity is observed to have increased, showing a stronger blueshift and a decrease in the line depth. Figure~\ref{fig:fig3} shows that this one-component inversion can sufficiently reproduce the observations.

For a portion of the filament there are multiple spectral components observed in the scan taken at 14:29 UT on May 30.  The less dominant components have usually a shallower line depth and large Doppler shifts. We will, however, not discuss  these features in more detail in the current study that focuses specifically on the properties of the erupting front of the filament.

\subsection{Fit of Beckers' Cloud Model (BCM) to \ac{halpha} 6562.8~\AA\ spectra}\label{SS:BCM}

\begin{figure}
\centerline{\includegraphics[width=1.0\textwidth]{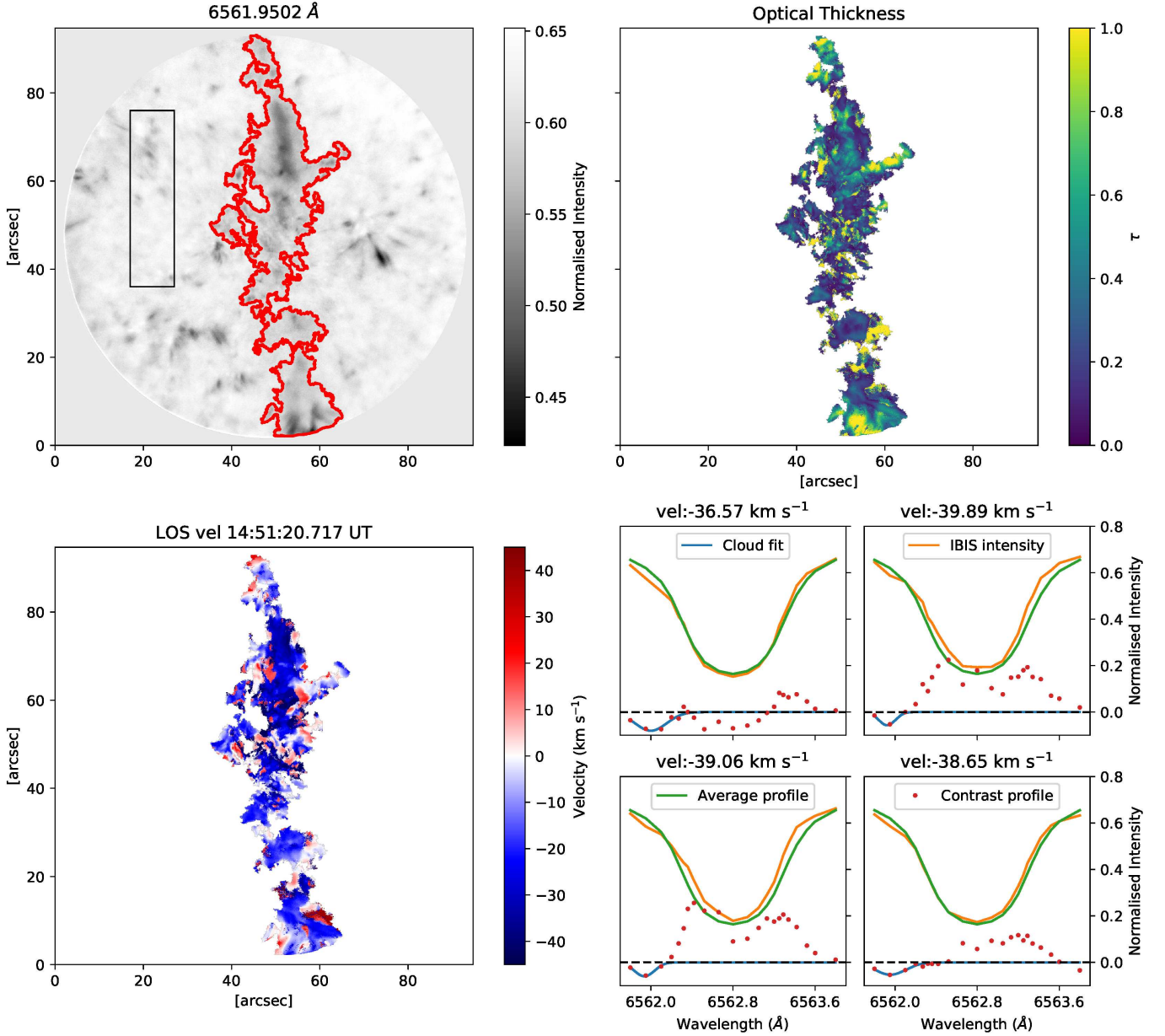}}
\caption{Method and results of the application of the  cloud model to the \ac{IBIS} \ac{halpha} observations. Top-left: normalised intensity at 6561.95~\AA\ wherein the absorption signature of the erupting filament is bordered by the red contour. Top-right \& bottom-left: optical thickness and \ac{LOS} velocity as derived from the BCM. Sample profiles are presented in the bottom-right panel. The green line indicates the average profile $I_0$ of \ac{halpha} derived as an average of all quiet-Sun profiles within the black box of the top-left panel. The orange line is the \ac{halpha} profile measured at a randomly chosen position within the \ac{FOV}. Red dots are the values of the contrast profile at the narrowband filter positions, and the blue line is the result of the BCM fitting to these points. The horizontal, dashed-black line highlights the level of zero normalised intensity.
\label{fig:BCM_inv}}
\end{figure}

We employed a cloud model following \cite{Beckers:1964} to fit the IBIS H$\alpha$ spectra. The simplifying approximations adopted by \citet{Beckers:1964} reduce the number of dependent variables of the \ac{RTE} to four; (1) constant background intensity $I_\mathrm{0}$ - The assumption that the background light incident across the studied pixel is constant, an assumption that may be less well-satisfied in more dynamic environments \textit{e.g.}, active regions, (2) constant source function $S$ - The assumption that the source function does not vary along the \ac{LOS}, (3) Gaussian-like optical thickness in wavelength - the assumption that the studied cloud is isotropic along the \ac{LOS}, and (4) a constant \ac{LOS} velocity \citep[][]{Maltby:1976,Raadu:1987,Kuckein:2016}. Each of these parameters may, of course, vary across the \ac{FOV}. As such, the \ac{RTE} reduces to the form,
\begin{align}
    I(\lambda)&=I_{0}(\lambda) \mathrm{e}^{-\tau(\lambda)}+S\left(1-\mathrm{e}^{-\tau(\lambda)}\right), \label{eq_beckers_rte}\\
        \tau(\lambda)&=\tau_{0} \mathrm{e}^{-\left(\frac{\lambda_{0}-\lambda}{\lambda_{\mathrm{D}}}\right)^{2}},
    \label{eq_tau}
\end{align}
where $\tau_{0}$ is the optical thickness of the line centre (assumed constant), $\lambda$ and $\lambda_{0}$ are the measured and rest wavelength \citep[calibrated as in][\textit{i.e.}, quiet-Sun spectral averages]{Schwartz:2019}, respectively, and $\lambda_\mathrm{D}$ is the total (thermal + non-thermal) Doppler width,
\begin{equation}
	\lambda_{\mathrm{D}}=\frac{\lambda_{0}}{c} \sqrt{\frac{2 k_\mathrm{B} T}{m} +\xi^{2}}, \label{eq_doppler_width}
\end{equation}
with $k_\mathrm{B}$ Boltzmann's constant, $T$ and $m$ the temperature and mass of the cloud, respectively, and $\xi$ the non-thermal velocity (NTV). In practice, and owed to the limited constraints on the components when using a single spectral line, we only solve for the total value of  the Doppler width and hence Equation~ (\ref{eq_doppler_width}) does not explicitly feature within the  \ac{BCM} method. \citet{Beckers:1964} introduced a further, seemingly arbitrary, simplification of Equation~ (\ref{eq_beckers_rte}) to,
\begin{equation}
    C(\lambda) \equiv \frac{I(\lambda)-I_{0}(\lambda)}{I_{0}(\lambda)}=\left(\frac{S}{I_{0}(\lambda)}-1\right)[1-\exp (-\tau(\lambda))], \label{eq_contrast}
\end{equation}
referred to as the so-called \textit{contrast profile}.

We used a constant source function and Doppler velocity, while the profile $I_\mathrm{0}$ (green spectra in the lower right panels of Figure \ref{fig:BCM_inv}) was derived from a quiet-Sun region within the \ac{IBIS} \ac{FOV} but away from the filament (black rectangle in the upper left panel of Figure \ref{fig:BCM_inv}). The source function, optical thickness, velocity, and line width model parameters were permitted to vary across the \ac{FOV} within the bounds [0.01,0.4] W m$^{-2}$ sr$^{-1}$ Hz$^{-1}$, [0,3], [-45,45] km s$^{-1}$, and [0.09,0.71]~\AA , respectively \citep[see \textit{e.g.},][]{Alissandrakis:1990, Chae:2006, Kuckein:2016}. The inversion procedure initially centers on the deepest portion of the profile to the blue side of the rest wavelength before solving the contrast equation (\ref{eq_contrast}) using the common iterative Levenberg-Marquardt least-squares fitting algorithm \citep[][]{levenberg:1944, marquardt:1963} implemented in the \acl{IDL}. Examples of the fitting results are shown in the bottom-right of Figure~\ref{fig:BCM_inv}. 

We restricted the cloud model fit to only the filament area. We defined a mask based on the average residual between normalised intensity and average profile $I(\lambda)-I_\mathrm{0}(\lambda)$
at 6561.8, 6561.95, and 6562.1~\AA\ with a 3-2-1 weighting. For each snapshot in time between 13:47\,--\,15:00~UT the intensity value for the contour varied in line with atmospheric seeing between -0.008 and -0.014. The resulting mask contour was then visually inspected to ensure that it did not include regions clearly not associated with the filament spine. We then extracted the average of the velocities within the mask that were both negative and had a corresponding optical thickness of less than 0.5 as a measure of the filament speed according to the \ac{halpha} observations. Regions within the mask that contained optical thicknesses $>0.5$ corresponded to much smaller, or even zero velocity, and  were often located at the outer boundary of the filament (upper right panel of Figure \ref{fig:BCM_inv}). This is suggestive of either a complicated internal structure/evolution within the erupting filament or that these regions, although isolated within the mask, may be more related to the properties of the background, low-altitude structures  rather than those of the filament body.

\subsection{Derivation of Temperature Estimates from He~{\sc i}~10830~{\AA}}
From Equation (\ref{eq_doppler_width}) one can derive estimates of the temperature $T$ and the NTV $\xi$ when simultaneous observations of two spectral lines from chemical elements with a significantly different molecular weight $m_i$ are available \citep[e.g.,][]{Bendlin:1988,2016SoPh..291.2281B}. The approach is valid for spectral lines that form in an optically thin medium. The roughly Gaussian shape of the He~{\sc i}~10830~\AA\ spectra (Figure \ref{fig:fig3}) and the optical depth in the HAZEL inversion results (see Section \ref{Sec:results_velo} below) support this for the He line. As we cannot reliably confirm it for the H$\alpha$ spectra in an on-disk filament observation and because the molecular weights of helium and hydrogen are rather close, we only used the full-width at half maximum (FWHM) of the He I 10830~\AA\ spectra to estimate the temperature in the filament material at a given NTV. 

To get estimates for the magnitude of the NTV in a reasonable range, we evaluated Equation (\ref{eq_doppler_width}) with the average FWHM of the He~{\sc i} line on May 29 prior to eruption of 0.75\,{\AA} by assuming average temperatures in the filament of $T=6, 10,$ and $20$\,kK and solving for the corresponding values of the NTV. This yielded three possible values $\xi_i =11.3, 10.5$ and $8.3$\,km\,s$^{-1}$, respectively. With those NTVs, we then converted the FWHM of the individual He~{\sc i} spectra on each pixel to three temperature estimates $T_i(x,y)$ and calculated average, minimal and maximal temperatures within the filament for the three different NTVs in each of the four FIRS maps.
\subsection{ChroTel Dopplergrams}
The ChroTel He I 10830~\AA\ observations covered the filament eruption from 07:15~UT to 17:09~UT on May 30, 2017. However, after 16:00~UT, the filament line depth decayed rapidly until the absorption signature was no longer present at around 16:30~UT. The Dopplergrams of ChroTel observations on May 30, 2017 were derived using the center-of-mass method described in \citet[][BE11]{Bethge:2011}  with some modifications. The median filtergram intensities $I_i$ for the filters $i=1...6$ were normalized to the median intensity in the seventh filtergram centered at 10833.15\,{\AA} that is least affected by solar spectral lines (see Figure 4 of BE11). The median value in a square covering about 6\,\% of the solar disk around disk center was calculated in each filtergram. Each filtergram was then multiplied by the ratio of the median intensity of the seventh to the actual filtergram:
\begin{equation}
 \tilde I_i=\frac{\overline{I_7}}{\overline{I_i}}\,I_i \,\mbox{\,\,for $i=1...6$}.
\end{equation}

The line-shift maps were then derived according to
\begin{equation}
 \Delta\lambda=\frac{\sum_i(I_7-\tilde
   I_{i})\lambda_{i}}{\sum_i(I_7-\tilde I_{i})}-\lambda_0 , \label{eq_chrotel}
\end{equation}

\noindent  where $\lambda_0$ is the rest wavelength of the He~{\sc i} line at 10830~{\AA}, which corresponds to an equal weight for each filter position ($\alpha_j \equiv 1$ in Equation 3 of BE11). The Doppler velocities are then derived as $v=(\Delta\lambda/\lambda_0)\cdot c$. 

The filament appeared in five Lyot filtergrams with the range of center wavelength [-2.7~\AA, +0.7~\AA] ($i = 1...5$) during the eruption.  For each observation, the filtergrams used to reconstruct the line shift value were selected  dynamically based on their signal strength, i.e. filtergrams  without a recognizable filament shape were rejected  and not included in the calculation of Equation  (\ref{eq_chrotel}).  Blue shifts of up to $-50$\,km\,s$^{-1}$ (filter $i=2$) are deemed reliable, but  not the values beyond that result from filter position 1 ($\equiv -80$\,km\,s$^{-1}$) that is strongly affected by the presence of the photospheric Si~{\sc i} line at 10827~{\AA} \citep[e.g., BE11,][]{Kuckein:2020}.

\subsection{Velocity Derivation from Imaging Data}

\begin{figure}
\centerline{\includegraphics[width=1.0\textwidth]{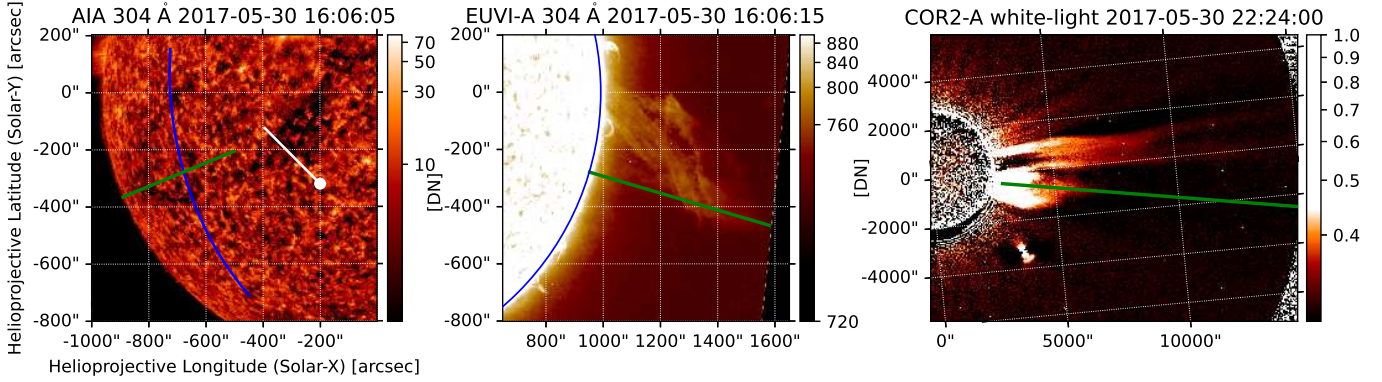}}
\caption{Left panel: map of SDO/AIA 304~\AA. Middle panel: STEREO-A/EUVI 304~\AA. Blue curves show the solar limb observed from STEREO-A. Right panel: STEREO-A COR2 white-light image. Green and white lines represent the position of slices.  An animation of the left and middle panels of this figure is available in the online material. It shows maps of SDO/AIA 304~\AA\ and STEREO-A/EUVI 304~\AA\ during the filament eruption from 00:06 to 18:56 UT on 2017 May 30. In both panels of the animation, white solid curves show the solar limb observed from STEREO-A.
\label{fig:fig5}}
\end{figure}

Example \ac{AIA}~304~\AA, \ac{EUVI}-A~304~\AA, and \ac{COR2}~white light imaging data that captured the eruption of the studied filament are shown in Figure~\ref{fig:fig5}. For each instrument, one slice was selected in the direction from the disk center to the filament front to construct the time slices shown later in Figure~\ref{fig:fig6}. For the AIA observations, the position where the filament first appears above the limb was used to set the slice direction. For both EUVI and COR2 observations, the feature point that is farthest away from the disc center was used to set the slice direction. The filament front positions were determined by a point-and-click method in the time slice of AIA 304~\AA, and were automatically selected based on the gradient along the slice with manual correction for points before 12:00~UT in the time slice of EUVI-A 304~\AA. The front positions of the CME in the STEREO-A COR2 white-light data were automatically selected based on the gradient along the slice. A Savitzky-Golay filter was applied to all three observations to smooth the position results \citep{2013A&A...557A..96B}. The uncertainties are estimated to be two pixels for both the \ac{AIA} and the \ac{EUVI} data. The velocities in the plane of sky were then obtained from the spatial derivative of the smoothed distance with time.

\section{Results} \label{sec:res}

\subsection{Line Width and Temperature}
\begin{figure}
\centerline{\includegraphics[width=1.0\textwidth]{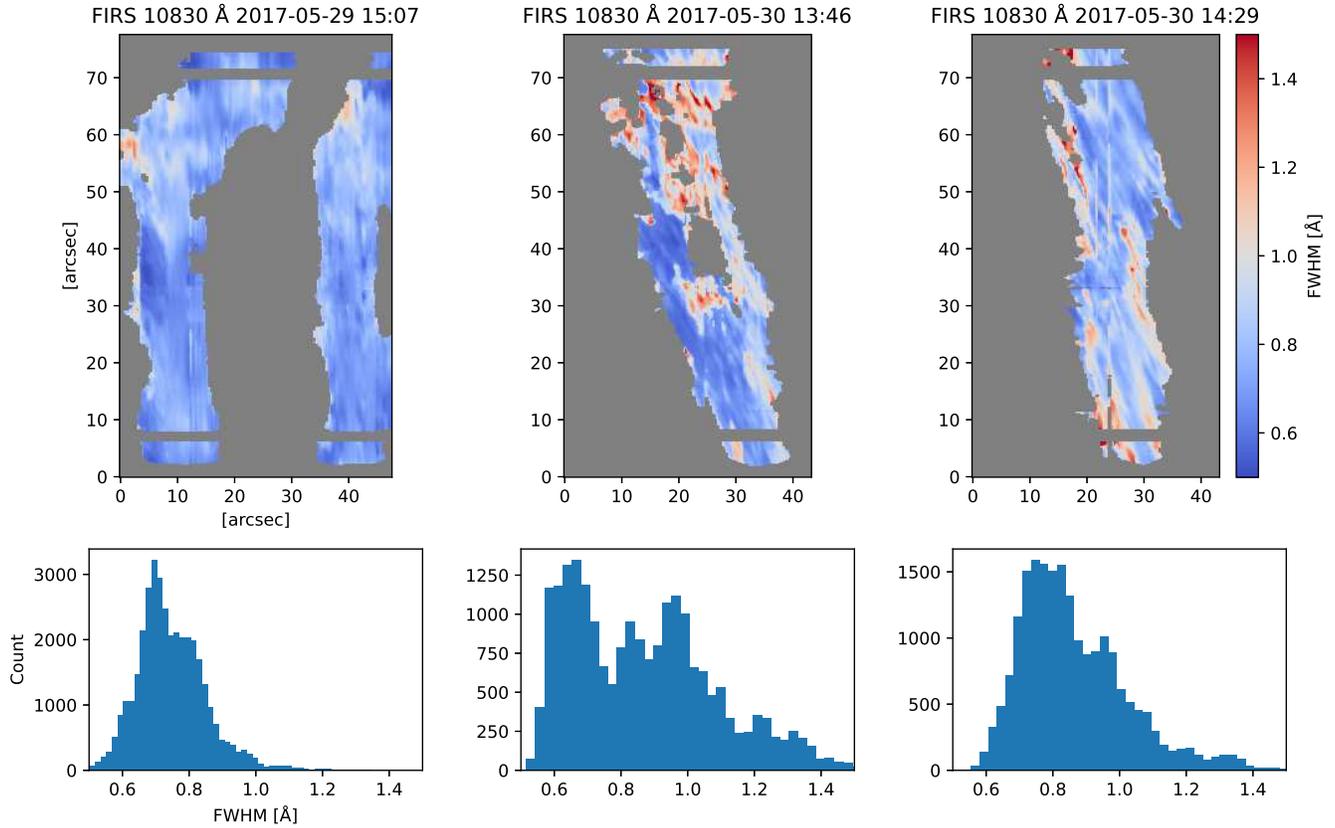}}
\caption{FWHM maps (top row) and histograms (bottom row) of the FIRS observations at He I 10830~\AA.  The top left panel shows the same portion of the filament observed twice on May 29 with a starting time of 14:41~UT at x = 0, and 15:07~UT at x = 33 arcsec, while the right two panels show the maps on May 30 at 13:46~UT and 14:29~UT.}
\label{fig:fig10}
\end{figure}

\begin{table}
\caption{FWHM and Temperature Values} \label{tab:1}
\begin{tabular}{| c| c c c | c c c | c c c |c c c |}
\hline
\multicolumn{4}{|c|}{T [kK] / $\xi$ [km/s]} & \multicolumn{3}{c|}{6 / 11.3} & \multicolumn{3}{c|}{10 / 10.5} & \multicolumn{3}{c|}{20 / 8.3}\\
\hline
Day & \multicolumn{3}{c|} {FWHM [\AA]} & \multicolumn{3}{c|} {T [kK]} & \multicolumn{3}{c|} {T [kK]}& \multicolumn{3}{c|} {T [kK]}\\  
 &  min & ave & max & min & ave & max & min & ave & max & min & ave & max \\  
\hline
29 & 0.55 & 0.75 & 1.07 & - & 6  & 46  & - & 10 & 50 & 4 & 20 & 60  \\
30 & 0.56 & 0.86 & 1.53 & - & 19 & 125 & - & 23 & 129  & 4 & 33 & 139 \\
\hline
\end{tabular}
\end{table}

Figure~\ref{fig:fig10} shows maps of the FWHM for the He I 10830~\AA\ spectra and the corresponding histograms. On May 29, the average line width was 0.75~\AA\ with a range from 0.55 to 1.07~\AA\ (Table \ref{tab:1}). On May 30, the mean value increased by about 15\,\% to 0.86~\AA, while the maximum value increased by up to 50\,\% to 1.53~{\AA}.  Using the three values of $\xi_i = 11.3, 10.5$ and $8.3$\,km\,s$^{-1}$, the maximal derived temperatures on May 29 were 46--60 kK and increased to 125--139 kK on May 30. The average temperatures found on May 30 are 13 kK higher than on the previous day independent of which  value of $\xi$ is used. For $\xi > 8.3$\,km\,s$^{-1}$, the minimum FWHM of 0.56~\AA\  on both May 29 and 30 would have to correspond to negative temperatures.

 From this simple estimate of temperature, an average temperature of about 20\,kK on May 29 and a mean increase by 13\,kK on May 30  with a non-thermal velocity below 8.3\,km\,s$^{-1}$ are the most likely results, while several small-scale areas forming elongated separate threads show significantly higher temperatures on May 30 (rightmost  top panel of Figure \ref{fig:fig10}).  Spectra with a large line width often only have a small line depth and show asymmetric line profiles with extended red wings. The inclusion of the extended red wing led to spurious large FWHM values for some of these spectra. The maximal derived temperatures are thus less reliable than the average values.

\subsection{LOS Velocities and Optical Depths from He~{\sc i} and \ac{halpha} Spectra} \label{Sec:results_velo}
\begin{figure}
\centerline{\includegraphics[width=0.75\textwidth]{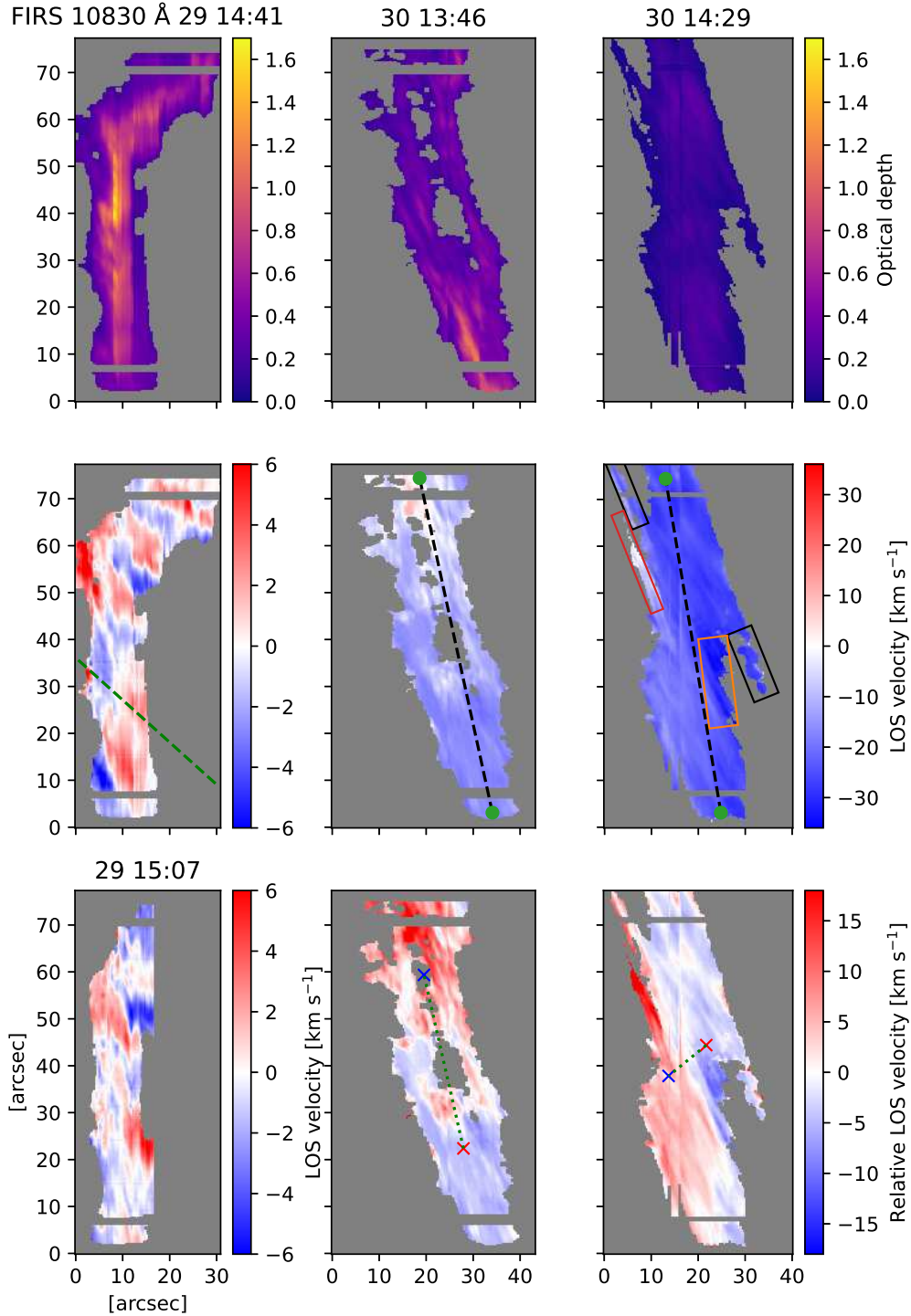}}
\caption{ Optical depth and line-of-sight velocity maps of the filament observed by FIRS at 10830~\AA.  Left to right: on May 29, May 30 at 13:46~UT and 14:29~UT. Top row: optical depth. Only the map at 14:41~UT is shown for May 29. Both observations on May 30 share the color bar at the right. First column,  bottom two panels: LOS velocities for the two maps on 29 May 2017 at 14:41~UT and 15:07~UT.  Middle row, right two panels: LOS velocities on May 30 at 13:46~UT and 14:29~UT.  Bottom row, right two panels:  LOS velocity on May 30 at 13:46~UT and 14:29~UT relative to the mean value of the whole filament within the FOV  in each map.  The green dashed line in the middle left panel shows the average magnetic field azimuth from Paper I. In the middle rightmost panel, two protrusions are marked with black rectangles, while red/orange rectangles indicate threads with small/large velocities. Dashed-black lines with green dots mark the positions of the filament axis used in Figure~\ref{fig:fig8}. In the middle and right panels of the bottom row, the centroids of blue/red shifted regions are marked as crosses in reversed color. They are connected with green dotted lines to show the direction of the gradient. Uniform gray pixels were not inverted.}
\label{fig:fig2}
\end{figure}

\subsubsection{He~{\sc i} 10830\,{\AA} LOS Velocities}
In Figure~\ref{fig:fig2} we present the \ac{LOS} velocity maps from the \ac{HAZEL} inversion. On May 29, the filament was stable as summarised in Paper I. The pattern in the LOS velocity map consists of elongated patches with widths of about five Mm that have their long axis parallel to the magnetic field lines whose directions have been provided already in Paper I. The average magnetic field direction forms an acute angle to the filament axis and is indicated by a green dashed line in the left middle panel of Figure \ref{fig:fig2}. Adjacent patches tend to have oppositely directed velocities. The two observations on this day have a time difference of 26 min at the same location. Despite being described as globally stable, the LOS velocity values may of course vary slightly or even reverse sign at any given local position. The general patterns remain, nonetheless, similar and may be explained as counter streaming flows along magnetic field lines with changing speed, oscillations perpendicular to magnetic field lines, or perhaps even signatures of individual magneto-thermal convection events. The fact that there are always white regions ($v\approx0$) between red and blue patches indicates velocity changes at the border of the patches with a smooth continuous transition; there is no imposition of lateral-atmosphere, pixel-to-pixel, coherency within the \ac{HAZEL} inversion tool. Along the direction of the magnetic fields there appears to be no change of sign in the LOS velocities apart from a few assumed threads  where the velocity changes sign. This may indicate a slight curvature of the magnetic field line relative to the LOS direction, as would be expected for the concave-up topology present within a magnetically-dipped portion of a flux rope. 

In the first observation at 13:46~UT on May 30, the filament was exhibiting blue shifts of about -11.0 km~s$^{-1}$ across its entire area. Most convincingly, the relationship between the velocity structure and the thin, elongated individual threads is now as apparent and clearly visible as in the  corresponding line-core intensity map of Figure~\ref{fig:fig9}. The filament width at this time was 20 arcsec, twice the width as on May 29. Maps of the \ac{LOS} velocity magnitude relative to the mean value of the filament within the FOV (middle bottom panel in Figure~\ref{fig:fig2}) show that the rising speed varies along the filament axis, with the south-east end rising faster than the other end  at 13:46~UT on May 30.

For the second observation at 14:29~UT, the mean LOS velocity of the filament has increased to -22.9 km~s$^{-1}$ with yet further distinctive, elongated threads than earlier. The region east of the filament spine marked with the red rectangle in the  middle-rightmost panel of Figure \ref{fig:fig2} shows nearly zero velocities. The two regions marked with black rectangles in the same panel show a small line depth and a LOS velocity of about -23 km~s$^{-1}$. The threads with the highest LOS velocities of -36.2 km~s$^{-1}$ are found around the region marked with an orange rectangle on the south-west side of the filament axis. The LOS velocities relative to the mean value of the filament within the FOV show that the rising speed is different perpendicular to the filament axis, with the north-west edge rising fastest  at 14:29~UT on May 30.

\begin{figure}
\centerline{\includegraphics[width=0.9\textwidth]{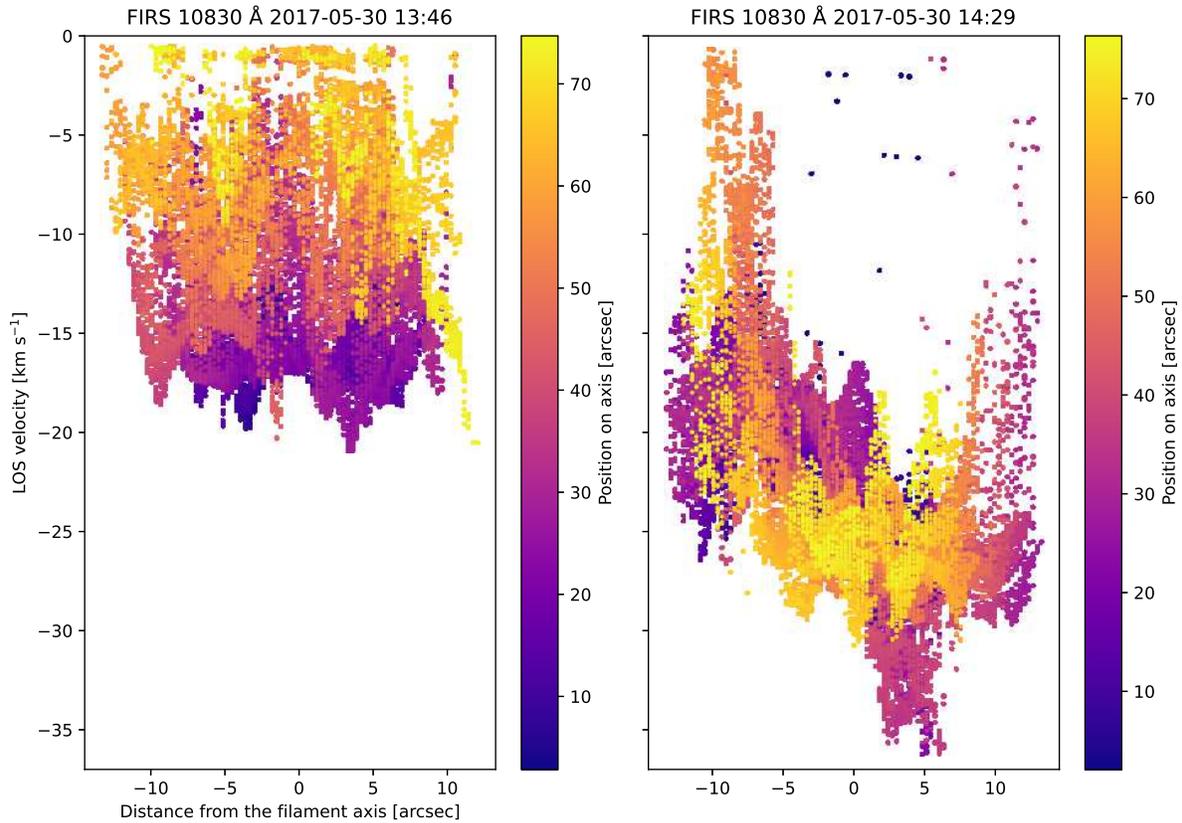}}
\caption{Scatter  plots of velocity and distance from the filament axis marked in Figure~\ref{fig:fig2}  on May 30 at 13:46~UT (left panel) and 14:29~UT (right panel). The zero  positions along the axis  are  marked with green  dots at the bottom of the axis line in each \ac{FOV} in Figure~\ref{fig:fig2}.
\label{fig:fig8}}
\end{figure}

In Figure~\ref{fig:fig8} the distribution and evolution of velocities at different distances from the filament axis on May 30 is further highlighted. The axis was determined by connecting two points which are centroids of the ends of the filament within the \ac{FOV}. These two points are shown in green at the top and bottom of the \ac{LOS} velocity panels of Figure~\ref{fig:fig2}. The velocity at 13:46 UT appears to have been symmetric to the filament axis, whereas the velocity at 14:29~UT was asymmetric with increasing values along the positive direction of distance from the filament axis. For the observation at 13:46~UT, the velocities were in the range of -5 to -18 km~s$^{-1}$. For the observation at 14:29~UT, the mean velocity on the left (right) of the axis was -20.4 (-26.2) km~s$^{-1}$. As such, there is a clear increase in the average velocity with time of 5.8 km~s$^{-1}$.  The two outer edges of the filament are about 20 arcseconds apart and were observed with a time difference of four minutes because of the sequential spatial scanning. Assuming a constant acceleration of 3.6 m s$^{-2}$ during the observation (the acceleration is obtained from the second derivative of the fitting line in Figure~\ref{fig:fig7}), a velocity difference of 0.8\,km\,s$^{-1}$ would be explained. The remaining difference of 5.0 km s$^{-1}$ between the right and left half of the filament indicates that the velocity distribution in the direction perpendicular to filament axis is skewed during the observation. This is explored in more detail in the next section. At 14:29~UT, the velocity distribution at both edges of the filament, where the coloured boxes of Figure~\ref{fig:fig2} were previously located, is broader than the central part.
\begin{figure}
\includegraphics[draft=false,width=0.85\textwidth]{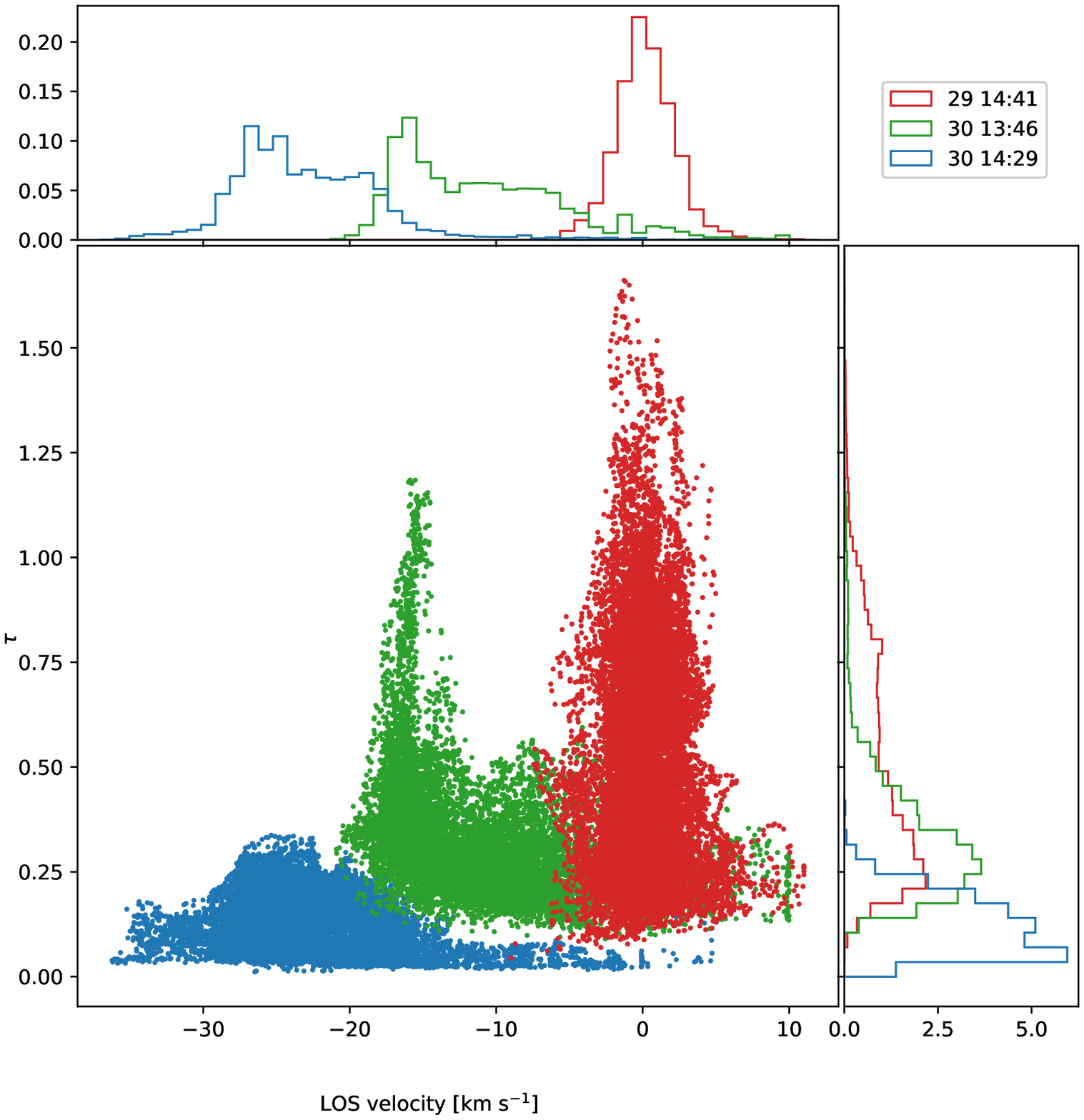}
\caption{Scatter plot of optical depth and LOS velocities with the corresponding histograms.  Red dots and histograms: on May 29 at 14:41~UT. Green (blue) dots and histograms: on May 30 at 13:46~UT (14:29~UT). The ROI used is marked in Figure~\ref{fig:fig9}.
\label{fig:fig11}}
\end{figure}

Histograms of LOS velocities for the three scans are shown in the upper panel of Figure~\ref{fig:fig11}. The width of the distributions increases significantly during the eruption in comparison with the pre-eruptive state. The mean value of LOS velocities is close to 0 on May 29, while on May 30, the values are -11.0 and -22.9 km s$^{-1}$, respectively. To describe the range of physical parameters in each observation, the range is defined as the difference between the 95th percentile and 5th percentile. The  ranges of the LOS velocities for the three observations  are $-6.5, -17.9$, and $-14.5$\,km\,s$^{-1}$, respectively.

The maps of the optical depth are shown in the top row of Figure \ref{fig:fig2}, while the distributions of optical depth are presented in the right panel of Figure~\ref{fig:fig11}. On May 29, the spine of the filament shows a continuous enhanced, relative to the background, optical depth along its full length with only a few short threads to the east  at about the middle of the FOV. The LOS velocity pattern has no discernible correlation with the optical depth. During the eruption on May 30, the optical depth maps show individual elongated threads that partially align with corresponding structures in the velocity maps. The optical depth values monotonically decreased during the eruption. The mean values of optical depth are 0.51, 0.34, 0.13 on May 29, at 14:41~UT, May 30 at 13:46~UT and 14:29~UT, respectively, with the ranges of optical depth also measured to have decreased, with values of 0.80, 0.46, 0.20. 

A scatter plot of optical depth and LOS velocities (lower left panel in Figure~\ref{fig:fig11}) shows that the data points of the three observations are separated from each other with small overlap. The filament was stable at 14:41~UT on May 29, with large widespread optical depths $\tau$ from about 0.1 up to 1.6 and LOS velocities around zero. During the eruption on May 30, the optical depth decreases to a maximum of $\tau=1.2$ at 13:48~UT and $\tau=0.3$ at 14:29~UT while the average velocities reach -11.0 and -22.9 km~s$^{-1}$, respectively.

\subsubsection{\ac{halpha} LOS Velocities}
In Figure~\ref{fig:BCM_inv} we present the  BCM inversion results for the \ac{halpha} spectra observed by the \ac{IBIS} instrument at the \ac{DST}. Velocities derived within the mask of the filament were primarily negative, \textit{i.e.}, towards the observer. Inspection of the fitting examples presented in the bottom-right of the Figure demonstrates that this is not imposed by the initial fitting procedure outlined in Section~\ref{SS:BCM}; the deepest portions of the observed profiles lie far into the blue wing of the \ac{halpha} profile. These plots also demonstrate that those profiles inverted within the absorption mask were generally shallower than the assumed-average profile within the \ac{FOV}, \textit{i.e.}, consistent with the weak absorption signature presented in the top-left intensity image of the same figure and Figure~\ref{fig:fig11}. Finally, the \ac{BCM} approach yields that the filament velocity increased from $\approx$~-10  to -22~km~s$^{-1}$ between 13:47~UT and 15:00~UT on May 30 for material with an optical thickness less than 0.5.

\subsubsection{ChroTel He~{\sc i}~10830~{\AA} Velocities}
\begin{figure}
\centerline{\includegraphics[width=1.0\textwidth]{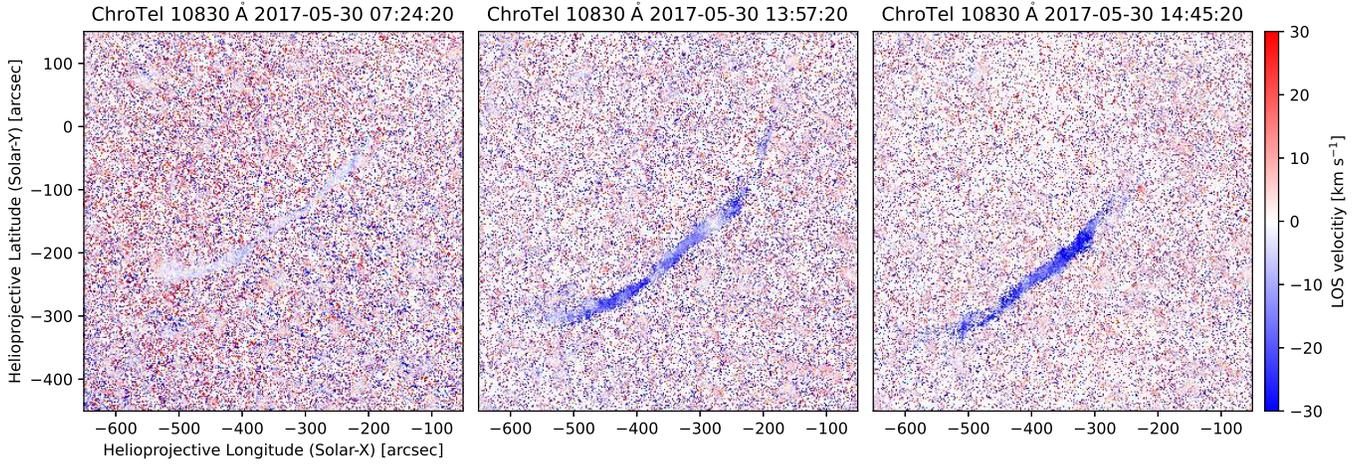}}
\caption{ChroTel He I 10830~\AA\ Dopplergrams on May 30, 2017. Quiet sun regions are noisy due to the shallow line depth. Left panel: the filament had LOS velocities close to zero at 07:24~UT. The filament is still  discernible due to its better signal-to-noise ratio than its surroundings. Middle panel: the whole filament was rising during the first FIRS observation on May 30 at 13:57~UT.  Right panel: the whole filament was rising faster during the second FIRS observation on May 30 at 14:45~UT.
\label{fig:fig12}}
\end{figure}

Finally, Figure~\ref{fig:fig12} shows three panels of the ChroTel Dopplergrams. The left panel is at the beginning of the ChroTel observations on May 30 at 07:15~UT with a mean LOS velocity of the filament of -5.3\,km\,s$^{-1}$. The two other panels were  obtained during the two FIRS scans on May 30 at 13:57~UT and 14:45~UT. The mean value of the LOS velocities are -14.7 and -19.5\,km\,s$^{-1}$, respectively. At 16:00 UT, the mean value of the LOS velocities reached -50.3\,km\,s$^{-1}$. After that the visibility of the filament gradually decreased.  The ChroTel observations thus provided a continuous, uninterrupted measure of the filament's speed at a 3 minute cadence from 7:15~UT until 16:00~UT that could be used to derive its acceleration.

\subsection{Evolution of the Filament Speed in the Plane of Sky}

\begin{figure}
\centerline{\includegraphics[width=1.0\textwidth]{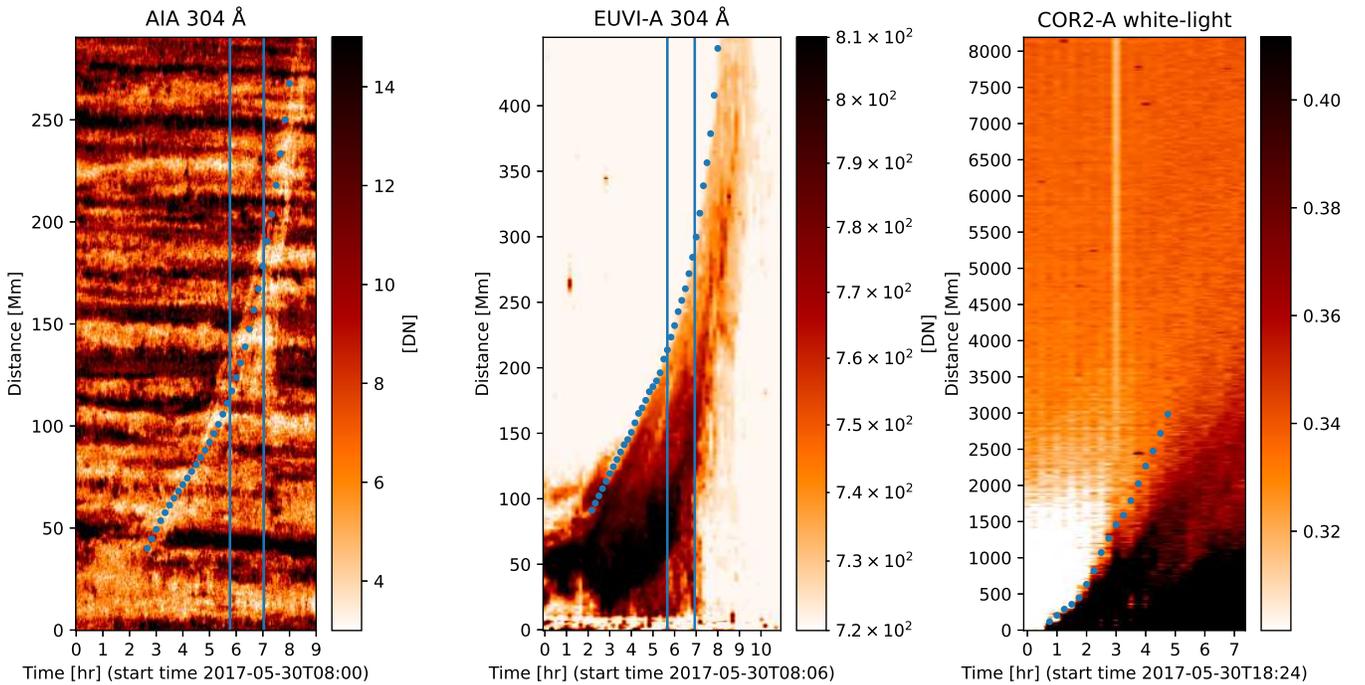}}
\caption{Left panel: Time slices of the SDO/AIA 304~\AA\ observations  on May 30. Middle panel: time slices of the STEREO-A/EUVI 304~\AA\ observations. Both 304~\AA\ observations are shown on an inverted colorscale.  The blue vertical lines indicate the time range of the DST observations. Right panel: time slices of the STEREO-A/COR2 white-light observations. Blue dots show the position of the filament/CME front.
\label{fig:fig6}}
\end{figure}

In addition to the \ac{FIRS} and \ac{IBIS} instruments the filament eruption was also observed by several other instruments which can be used to derive velocities; SDO/AIA and STEREO-A/EUVI at 304~\AA\ and STEREO-A COR2 in  white-light (see Figure~\ref{fig:fig5} and its associated online animation) . The results for the tracking of the leading edge of the filament as observed by the three instruments are summarised in the three panels of Figure~\ref{fig:fig6}.

The filament started moving at 10:30~UT  (corresponding to t = 2.5 hrs in the left two panels of Figure~\ref{fig:fig6}) according to the 304~\AA\ observations taken by the \ac{AIA}. By 16:10~UT on May 30, the projected filament front reached the solar limb, and the portions of the filament that project against the background of space are no longer visible  (see the animation in the online material). In the EUVI 304~\AA\, channel the filament front reached the edge of the field of view at 16:10~UT  (corresponding to t = 8 hrs in the left two panels of Figure~\ref{fig:fig6}) with a velocity of 60.1~km~s$^{-1}$. 

\begin{figure}
\centerline{\includegraphics[width=0.85\textwidth]{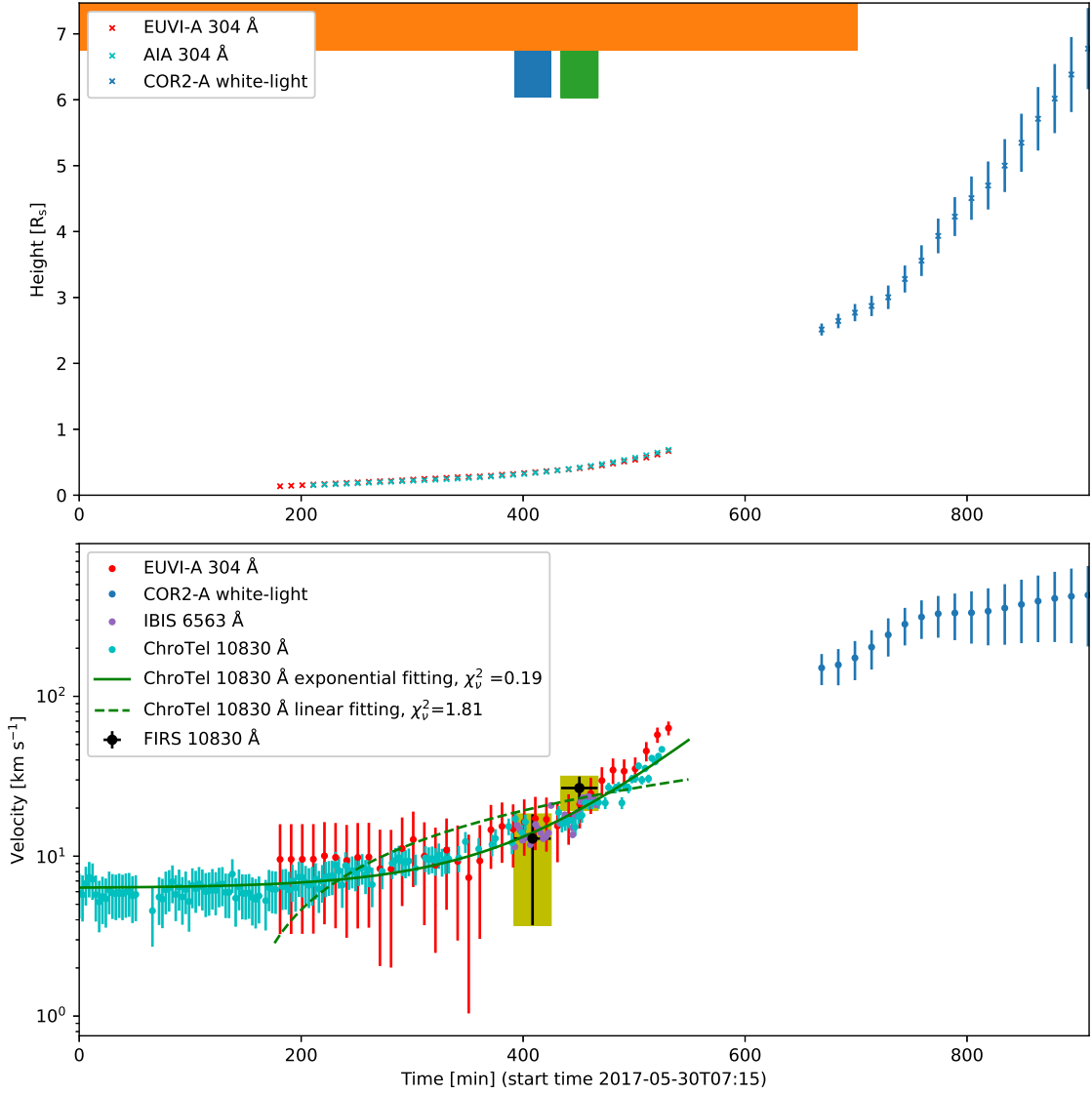}}
\caption{Upper panel: evolution of filament height. Error bars represent 3$\sigma$.  The orange rectangle at the top shows the time range of the animation that overlaps with the panel, while blue and green rectangles indicate the time range of the DST  observations. Lower panel: evolution of filament velocity. The black vertical bars representing FIRS velocity show 10th/90th percentile at lower/upper ends. The horizontal bars mark the observing time with median value of velocities. Error bars for EUVI-A, COR2-A, and ChroTel represent 1$\sigma$.
\label{fig:fig7}}
\end{figure}

 The STEREO-A COR2 observations show that the filament eruption was associated with a \ac{CME}. The \ac{CME} observed in STEREO-A COR2 white-light can be seen in the right panel of Figure~\ref{fig:fig5}. The direction of the slice in the COR2 white-light observation is the same as the direction of the slice observed from the STEREO-A/EUVI 304~\AA\ shown in the middle panel of Figure~\ref{fig:fig5}. The resulting time-slice image through the center of the \ac{CME} is shown in the right panel of Figure~\ref{fig:fig6}. As can be seen in Figure~\ref{fig:fig6}, both the EUVI and the COR2 white-light coronagraph observed a propagating intensity decrease which implies a density depletion of the associated \ac{CME} as it expands outward. Starting at 18:24 UT the \ac{CME} had a velocity of 145.6~km~s$^{-1}$ that increased to a final velocity of 430~km~s$^{-1}$ at 22:24 UT after which it became too faint in the COR2 coronagraph images (details of the \ac{CME} velocity can be found in Figure~\ref{fig:fig7}).

\subsection{Velocity Evolution of the Erupting Filament}

With the assumption that the direction of the filament eruption was radial, we derived the deprojected height of the filament/CME front (top panel of Figure~\ref{fig:fig7}). The heights derived from EUVI 304~\AA\ and AIA 304~\AA\ are consistent, and the COR2 instrument tracks a much later stage in the \ac{CME} evolution. For the velocity diagnostics, all LOS data were converted to a rising speed also in the radial direction. The velocity of the erupting filament according to EUVI and COR2 were derived from its height (see bottom panel of Figure~\ref{fig:fig7}) where the results are consistent across both instruments. Then, the continuous observations of ChroTel He~{\sc i}~10830~\AA\ are over an extended period of time that subsequently enabled us to fit both the early and late eruption phases.

The mean velocity  derived from the ChroTel He~{\sc i}~10830~\AA\ data, and both an exponential and a linear fit are shown in the lower panel of Figure~\ref{fig:fig7}, with a reduced $\chi^2$ of 0.19 and 1.81, respectively. Hence, the velocity curve during the eruption appears most-consistent with an exponential growth. The exponential fitting gives a value of 6.3~km~s$^{-1}$ for its horizontal asymptote. The uncertainty of the velocity of the ChroTel He~{\sc i}~10830~\AA\ data was estimated based on the difference between the observed and fitted values to be about 1.8~km~s$^{-1}$. The mean filament velocity reached 46.6~km~s$^{-1}$ on May 30 at 16:00~UT. The CME velocities reached 350 km~s$^{-1}$ on May 30 at 22:20~UT. The acceleration value was 1.7 (2.8) m~s$^{-2}$ at 14:05~UT (14:45~UT) when the first (last) FIRS observation on May 30 was halfway through, and subsequently increased to 7.4~m~s$^{-2}$ at 16:00~UT, when the filament was about to disappear in the 10830~\AA\ observation due to decreased line depth. The acceleration in the CME phase was derived from the COR2 heights to have been 12 m~s$^{-2}$ on May 30 at 21:20~UT.

\section{Discussion} \label{sec:dis}

\subsection{Derivation and Comparison of Filament  Velocities During Eruption} \label{subsec:filtrack}

\citet{Kuckein:2020} analyzed an eruption of part of a quiescent filament with blue-shifted line profiles exhibiting different shapes. They advocated convincingly for the use of k-means clustering to avoid inverting physically different spectra with a single model. However, we do not find any regions that show line profiles containing significant asymmetric line wings in this event. Of course, we already selected a subset of available profiles within the observations with the use of an inversion mask that isolated the deepest profiles believed to be related exclusively to the erupting filament. All line profiles of Stokes I for this filament observed by FIRS show one dominant component with symmetric line wings during the eruption. Many previous reports of events with \ac{LOS} velocities $>$~20 km~s$^{-1}$ observed in He~{\sc i}~10830~\AA\ spectra are also accompanied by a distinct component at rest ($<$ 8~km~s$^{-1}$) \citep[e.g.][]{1998ASPC..155..341M,2000ApJ...544..567S,Sasso:2011,Sasso:2014,2016ApJ...833....5S}. However, this erupting quiescent filament did not exhibit a component at rest. Crucially, the aforementioned papers studied targets predominantly within active regions and performed inversions across their entire \ac{FOV}, while the event presented here occurred within the quiet Sun and only specific regions of the \ac{FOV} were analysed. We therefore concur with a possible conclusion suggested before that the component at rest observed by these previous authors is likely associated with stronger photospheric magnetic field beneath the filament that is absent for this event \citep[cf.][]{Diazbaso:2016,Diazbaso:2019a,Diazbaso:2019b,Diazbaso:2019c}.  Another explanation could be that because the position of the filament is far from disk center, the inclined LOS does not scan the lower part of the filament but a quiet region far from the position of the eruption source.

It is worth noting that some threads in one region showed much lower rising speeds of 2--6 km~s$^{-1}$. Assuming that the plasma  has some average velocity during the eruption, and neglecting the possibility that this signature is sourced below the erupting filament, plasma at a \ac{LOS} velocity of about zero must correspond to downward flows along threads relative to the rising body of the flux rope. The location of these threads hints at a potential relation to a barb that previously connected the filament to the photosphere, although this is purely a spatial correlation \citep{Jenkins_thesis:2020}.

The fit for deriving the  velocity profile in Figure~\ref{fig:fig7}  was applied to the ChroTel He~{\sc i}~10830~\AA\ full-disk chromospheric Dopplergrams. Only synoptic full-disk instruments such as ChroTel or the Solar Flare Telescope \citep{2020JSWSC..10...41H} can currently provide the data needed for measuring \ac{LOS} velocities of on-disk eruptions that might be used to refine an estimated time of arrival for space weather prediction purposes in near real-time. Nevertheless, the rising speed of the erupting filament was derived from different data sources with consistent results. Numerical simulations show that an eruption driven by breakout reconnection exhibits a height profile best fit with a quadratic function \citep[\textit{e.g.},][]{2004ApJ...617..589L}, while a kink/torus instability requires an exponential function \citep[cf.][]{2019ApJ...882...85O}.  A quadratic function of the  height profile is often seen in prominence eruptions \citep[\textit{e.g.},][]{2003ApJ...586..562G,2015ASSL..415..381G,2020ApJ...894...85C}, which would correspond to a linear function fitted to the velocity profile in the lower panel of Figure~\ref{fig:fig7}. In this observation, the velocity profile derived using the He~{\sc i}~10830~\AA\ Dopplergrams is fitted well with an exponential function, thus the observed eruption is consistent with a kink/torus instability as the driving mechanism. 

The ChroTel data at 07:15~UT on May 30 indicate that the filament had been perturbed prior and had already gained a mean upward velocity of 6.3 km~s$^{-1}$ by that time. Unfortunately, the ChroTel data  do not extend further back in time than 07:15~UT, and the  high-resolution observations of either \ac{FIRS} or \ac{IBIS}  on May 29 preceded the initiation, as indicated by their observations of a \textit{stable} filament, by some 10 hours or so. As such, we are unable to suggest which of the many possible trigger mechanisms was responsible for the slow evolution preceding the eruption. There are some filament eruption events associated with CMEs and ICMEs reported with upward velocities observed hours before the eruptions, similar to the event that we have presented here \citep{2020JSWSC..10...41H}. Telescopes with off-band \ac{halpha}~6562.8~\AA\ or He~{\sc i}~10830~\AA\ capability are able to detect this kind of filament eruption through Dopplergrams about half a day before its motion shows up in chromospheric line-center observations. It is therefore unfortunate that in most cases the instrumentation at telescopes that have a synoptic program currently lack the ability to perform such observations.

The analysis of the high resolution \ac{halpha} observations of \ac{IBIS}  yields velocities that are consistent with those obtained from both \ac{FIRS} and ChroTel. Quantitatively, the velocity in \ac{halpha} is observed to have increased from rest to -10\,--\,-22~km~s$^{-1}$ during the period of observation. Furthermore, and most crucially, the results of all of the spectroscopically-derived velocities are in agreement, to at least the same order of magnitude or better, with those velocities derived using the imaging instruments of \ac{SDO}/\ac{AIA} and \ac{STEREO-A}/\ac{EUVI}.

The methods employed to extract the velocities from both the He~{\sc i} and \ac{halpha} observations may be considered simplistic in their handling of the radiative transfer theory. Nevertheless, authors such as \citet{Mein:1996} have shown that, for \ac{halpha}, the discrepancies between the results of the \ac{BCM} and a fully-\ac{NLTE} model may be of the order of only a few tens of \% and only critical for those filaments with an optical thickness much larger than one, \textit{i.e.}, larger than measured for the filament studied here. However, although a valid conclusion for comparatively stable filaments, such a relationship may become of second-order importance when considering eruptive geometries; the assumption of a 1D, plane-parallel atmosphere with zero lateral photon-loss will undoubtedly become increasingly invalid with increasing altitude and internal structural complexity. It is imperative to understand the finer details of conditions present within the filament plasma in general, however, it appears from Figure~\ref{fig:fig7} that the addition of such considerations \citep[\textit{e.g.},][]{Heinzel:1999b,Tziotziou:2007,Schwartz:2019} to the simple models used in this study are not necessary to extract complimentary information (agreement with other models to within a few km~s$^{-1}$) so as to consistently characterise the early velocity evolution within an erupting filament. Naturally, this does not exclude the consideration that each of the spectral inversion methods may be similarly incomplete. 

\subsection{Additional Points of Interest} \label{subsec:adddisc}

Our study here focuses primarily on the evolution of plasma velocity within an erupting quiescent filament, measured using a combination of spectroscopic and monochromatic observations and their associated analysis tools. Nevertheless, these tools also provide additional parameters, and the observations contain additional features, that we consider to be of interest to the wider community. 

To begin, the properties of the plasma within the pre-eruptive filament have previously been studied in detail by \citet{Schwartz:2019}, where the authors performed a careful analysis of the \ac{halpha} absorption at six positions along the filament spine. Although a less focused approach, the more-general \ac{HAZEL} tool has enabled us here to invert the entire \ac{FOV} and as such we have access to the spatial variation of the radial velocity on a scale of  about $\approx$~100\,--\,200~km. A general one-to-one, pixel-to-pixel comparison of these maps to the parameters inverted by \citet{Schwartz:2019} would require a separate, dedicated study. Nevertheless, the global, striped pattern in the radial velocity is intriguing for a number of reasons. Similar observations have previously been interpreted as signatures of counter-streaming material along the host magnetic field \citep[e.g.,][and many subsequent citations]{Zirker:1998}. In this case, we find coherent, $\approx$~5~Mm width plasma motions aligned with the azimuth field vector as deduced in paper I. The occasional reversal in sign of the motions may thus represent the projected velocity of material flowing coherently away from the observer on one side of the filament and towards on the other. The consideration of a flux rope topology, as deduced in paper I, then  points to the hypothesis that  filament material was flowing in different directions (counter-streaming) around the inside of a flux rope. The occasional reversal in sign along a given flux tube (cf. Paper I) thus illuminates the concave-up shape inherent to the magnetic  configuration \citep[cf. simulations of][]{Jenkins:2021}. Alternatively, assuming the motions of the plasma were oriented parallel to the \ac{LOS} they thus describe material flowing towards and away from lower heights. If so, such undular velocity patterns may be the filament counterpart of the magneto-thermal convection frequently recorded within quiescent prominences above the limb i.e., radial striations induced by the Rayleigh-Taylor instability \citep[ see][ and references therein]{Hillier:2018}. The closely-arranged red- and blue-shifted regions would thus correspond to the `falling fingers' and `rising plumes', respectively. However, the ability to confidently distinguish either behaviour from general small-scale oscillations would require a more detailed study that lies outside of the scope of the current work.

Returning to the eruptive phase, the \ac{FIRS} inversion results of the \ac{LOS} velocities on May 30 indicate that the filament motion may be decomposed into three categories: the erupting translational motion in the radial direction which has the largest magnitude, the flow motion along magnetic field lines that highlights the thread structure, but also a possible third, rotational motion about the main axial field. For material flowing around a cylinder, one would expect to observe a velocity gradient across the centre of the cylinder associated with a smooth variation in the alignment between the \ac{LOS} and the cylinder edge. Presuming that we may consider the magnetic `cage' in which the filament material is evolving to be both symmetric and translationally invariant along its axis, the cross-axis gradient signature is indeed suggested in the bottom-right panel of Figure~\ref{fig:fig2}. However, the relative velocity gradient (green dotted lines in the bottom row middle and right panels of Figure~\ref{fig:fig2}) is distinctly different between 13:46 and 14:29~UT on May 30, with the gradient clearly being along the axis, rather than across it, for the earlier scan. At  13:46~UT and for the position of the \ac{FOV}, the filament and its bounding magnetic field will have been both closer to the surface and had more curvature to its axial field. The gradient along the axis may be explained by this assumed curvature in the same direction \citep[cf.][]{Titov:1999,Xia:2016,Kaneko:2018}.

\begin{figure}
\centerline{\includegraphics[width=0.9\textwidth]{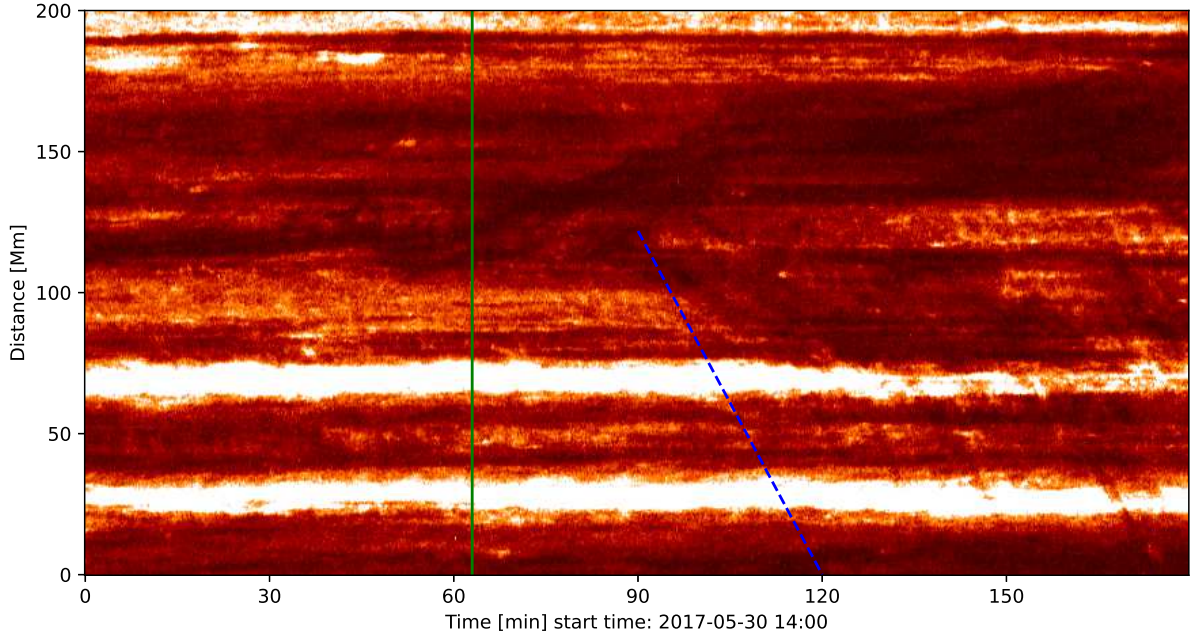}}
\caption{Time slices of the SDO/AIA 304~\AA\ observations. The slice position is shown as white line and a white dot marking the starting point in the left panel of Figure~\ref{fig:fig5}.  The green solid line marks the time of the end of the last FIRS observation at 15:02~UT. After 15:30~UT, there are a number of dark stripes that are parallel to the blue dashed line that may indicate the untwisting motion of the filament. The gradient of the blue dashed line is $\approx~17$~km~s$^{-1}$.}
\label{fig:fig17}
\end{figure}

The position of the \ac{FIRS} \ac{FOV} changed between the two scans in an attempt to follow the erupting structure and as such there is no guarantee that the two scans observed the same portion of the erupting structure. Consider, first, the possibility that the tracking was successful and the regions observed by the two snapshots are related. The observed expansion would presumably involve a straightening of the legs of the erupting structure, an evolution in the gradient of the velocity along the same portion of the filament axis would then be expected. This would not, however, necessarily explain the shift in the gradient direction from along to across the axis. Such a shift would require either a sudden and significant flow along the assumed-helical magnetic field, or a rotation of the magnetic field around the axis itself. In the absence of a reasonable hypothesis for such a sudden and bulk flow of plasma, we speculate that it is instead more likely that this change in gradient orientation is a consequence of an unravelling motion associated with the expanding magnetic field. The untwisting of filaments  and prominences during eruption has previously been reported  by, e.g.,  \citet{2012A&A...540A.127K}, \citet{2016NatCo...711837X} and \citet{ Kuckein:2020}. Figure~\ref{fig:fig17} presents a time slice of the AIA 304~\AA\ observations showing that the absorption signature of the filament started to expand around 15:00 UT on May 30 \textit{i.e.}, at the end of the last \ac{FIRS} scan. After 15:30 UT, there are many dark stripes that are  parallel and appear to show the motion of the filament threads \citep[cf. Figure 1 of][]{2016NatCo...711837X}. The gradient of the blue dashed line overlaid on this Figure equates to $\approx~17$~km~s$^{-1}$, slightly larger than the magnitude of the relative velocity on either side of the assumed-axis shown in Figure~\ref{fig:fig2}. Then, assuming the material captured in 304~\AA\ absorption is located in the underside of the flux rope, a common assumption for a stable filament, the extension of these `threads' towards the bottom of Figure~\ref{fig:fig17} is also consistent with the position of the (relative) red-shifted portion of the filament in the bottom-left of the bottom-right panel of Figure~\ref{fig:fig2}.  A second possibility would be that on May 30 the region captured in the second \ac{FIRS} snapshot at 14:29~UT was closer to the top of the erupting filament than the first observation at 13:46~UT. The apparent velocity gradient across the axis recorded in Figure~\ref{fig:fig2} may  then instead indicate the flow of material associated in some way with the expanding bow of the eruptive structure. Nevertheless, the independent observation of these moving `threads' in Figure~\ref{fig:fig17} remains, wherein the associated cut (white line) in Figure~\ref{fig:fig5} is positioned across the western leg of the erupting structure. Unfortunately, we are unable to confidently distinguish between these two possibilities without additional information.

Figure~\ref{fig:fig11} shows how the optical depth decreased during the eruption, while both the velocity  (Figure~\ref{fig:fig7}) and the lateral width (Figure~\ref{fig:fig9}) of the filament increased instead. In addition, we estimated a consistent temperature increase of 13 kK between the quiescent and rising phase in  Table~\ref{tab:1} for  non-thermal velocities between 8 and 11 km\,s$^{-1}$ with a most likely average pre-eruptive temperature of 20\,kK. In some regions of their erupting prominence, \citet{Zhang:2019} found NTVs below 9~km~s$^{-1}$ along with a smaller temperature increase of a few hundred K during the activation phase. Observational determinations of temperature and NTV in limb spicules range from 6--20\,kK and  5--24\,km\,s$^{-1}$ \citep{Bendlin:1988,socasnavarro:2005,2016SoPh..291.2281B,2018SoPh..293...20A},  where \citet{2016SoPh..291.2281B} found up to 50\,kK in a macrospicule. For filaments, a temperature range of 10\,kK for the core and up to 200\,kK for the \textit{prominence-corona-transition-region} layer, \textit{i.e.}, the outer boundary of a filament thread, has previously been reported \citep[][]{Labrosse:2010,Parenti:2014,Vial:2015}. The microturbulent velocity within prominences has commonly been assumed to be approximately 5~km~s$^{-1}$ with only a limited number of corroborating observational studies \citep{Gouttebroze:1993,Tziotziou:2001,Schwartz:2019}, while \cite{2015A&A...582A.104R} found values $>$ 15 km\,s$^{-1}$ in an Ellerman bomb. The corresponding values in the current study thus align with previous findings. The increase in the average temperature suggests that an increased degree of ionization of Helium may be at least partly responsible for the reduction of the opacity, while the aforementioned expansion (lateral and symmetric or involving an untwisting) would contribute to the same effect by spreading the mass contained in the filament over a larger volume.

 From the perspective of AIA, the filament reached the solar limb around 16:00 UT on May 30, 2017. The AIA 304 \AA\ observations show that the filament disappeared once projected above the limb (see the animation). This is different from stable filaments which often appear clearly as prominences when rotated above the limb. The absorption signature of this eruptive filament had a mean intensity of 1.3 DN as it approached the limb. According to the upper panel of Figure~\ref{fig:fig7}, the filament was at a height of about 500 Mm at this time. Assuming the absorbed light was subsequently re-emitted isotropically, the dilution factor takes a value of 0.094  \citep[see 5.4.2.2 of][]{
    2015ASSL..415..103H}. As such, the mean intensity of the filament once it rises above the limb and transitions to a prominence is expected to be a maximum of 0.13 DN. This expected value is an order of magnitude lower than the AIA 304 \AA\ read noise \citep[see Table 6 of][]{2012SoPh..275...41B}, and so it is not surprising that the prominence signal is not detected in the AIA 304 \AA\ data.

Finally, it is of interest that there was a small coronal hole (CH) close to the disk center visible in the AIA 193~\AA\ data on May 30. CHs are the source of high-speed streams (HSS) in the solar wind. The solar wind speed observed by the Advanced Composition Explorer showed an increase of wind speed on June 3, going up to around 500 km~s$^{-1}$. This would correspond to a transit time of around 3.5 days, compatible with the CH close to disk center on May 30. Most of the in-situ solar wind characteristics of this event between June 3 to June 5 is that of a HSS. Nevertheless, the magnetic field data from 11:00~UT to 18:00~UT on June 3 indicate its components are smooth and switch sign, which is not usually the case in a HSS but typical for a magnetic flux rope. A possible explanation is that the western flank of the CME got embedded in the HSS originating from the small disk center CH and both the HSS and the CME flank arrived together on June 3, travelling closer to a speed of 500 km~s$^{-1}$.

\section{Conclusions} \label{sec:con}
We have derived the propagation velocity, in addition to a variety of additional parameters, for an erupting large-scale filament from a series of multi-instrument imaging and spectroscopic data. Importantly, we have successfully demonstrated consistency between the ejection velocity measured spectroscopically and the speed inferred using the propagation of filament material from monochromatic images. The velocity profile during the eruption is better reproduced by an exponential than a linear function. This behavior is in favour of a kink/torus instability, which requires a flux rope. The existence of a flux rope is consistent with the corresponding results concerning the magnetic topology found in Paper I \citep{Wang:2020}. We conclude that synoptic full-disk chromospheric Doppler measurements can provide a near real-time determination of the rise speed of on-disk erupting filaments which might be used in future data-driven \ac{CME} propagation models.

\vspace{5mm} 
\textit{Acknowledgements.} We wish to thank the anonymous referee for their constructive comments that helped with the clarity of the arguments presented, and generally improved the quality of the manuscript. This work was funded by NSF grant 1839306. Sunspot Solar Observatory is a multi institution consortium that is funded by multiple entities including NSF (1649052, 1945705) and the State of New Mexico. Funding for the DKIST Ambassadors program is provided by the National Solar Observatory, a facility of the National Science Foundation, operated under Cooperative Support Agreement number AST-1400405. J.M.J.~thanks the STFC for support via funding given in his PhD Studentship, travel funds awarded by the Royal Astronomical Society, and support given as a part of the ERC Advanced-grant PROMINENT. K.M. acknowledges support by the NASA Heliophysics Guest Investigator program, the NASA cooperative agreement NNG11PL10A and the ISFM program at NASA GSFC. D.M.L.~acknowledges support from the European Commission's H2020 Programme under the following Grant Agreements: GREST (no.~653982) and Pre-EST (no.~739500) as well as support from the Leverhulme Trust for an Early-Career Fellowship (ECF-2014-792) and is grateful to the Science Technology and Facilities Council for the award of an Ernest Rutherford Fellowship (ST/R003246/1). D.P.C.~was partially supported through NSF grant AGS-1413686.

\bibliography{bibliography}
\end{document}